\documentclass[aps,onecolumn,preprint,prd,superscriptaddress,nofootinbib]{revtex4}
\usepackage{amsmath,amssymb,color,mathrsfs, graphicx,verbatim,epsfig, bbm, wasysym, pstricks}

\newcommand{\beq}{\begin{equation}}
\newcommand{\eeq}{\end{equation}}
\newcommand{\beqa}{\begin{eqnarray}}
\newcommand{\eeqa}{\end{eqnarray}}
\newcommand{\ba}{\begin{array}}
\newcommand{\ea}{\end{array}}
\newcommand{\CO}{\mathcal{O}}

\begin{document}

\begin{titlepage}
 
\thispagestyle{empty}

\begin{flushright}UMD-PP-11-002\\ RUNHETC-2011-06
\end{flushright}
\vspace{0.2cm}
\begin{center} 
\vskip .5cm
{\Large \bf   (De)Constructing a Natural and Flavorful Supersymmetric Standard Model}
\vskip 1.0cm

\vskip 1.0cm
{\large Nathaniel Craig,$^{1,2}$ Daniel Green,$^{1}$ and Andrey Katz$^3$}
\vskip 0.4cm
{\it $^1$School of Natural Sciences, Institute for Advanced Study,  Princeton, NJ 08540}
{\it $^2$Department of Physics and Astronomy, Rutgers University, Piscataway, NJ 08854}
{\it $^3$Department of Physics, University of Maryland, College Park, MD 20742}
\vskip 1.7cm
\end{center}

\noindent 
Using the framework of deconstruction, we construct simple, weakly-coupled supersymmetric models that explain the Standard Model flavor hierarchy and produce a flavorful soft spectrum compatible with precision limits. Electroweak symmetry breaking is fully natural; the $\mu$-term is  dynamically generated with no $B \mu$-problem and the Higgs mass is easily raised above LEP limits without reliance on large radiative corrections. These models possess the distinctive spectrum of superpartners characteristic of ``effective supersymmetry": the third generation superpartners tend to be light, while the rest of the scalars are heavy.

\end{titlepage}

\section{Introduction}

Weak-scale supersymmetry (SUSY) provides a compelling solution to the Standard Model (SM) hierarchy problem. However, collider limits on the Higgs and scalar masses have already led to some tension with the naturalness of weak-scale SUSY, while precision measurements have significantly constrained relations among SUSY-breaking parameters. 
In order to provide a truly satisfying solution to the hierarchy problem, a supersymmetric Standard Model must therefore fulfill a variety of additional requirements beyond the cancellation of quadratic divergences.  These include:
\begin{enumerate}
\item {\it Satisfaction of various precision constraints.}

The mechanism of SUSY-breaking should not introduce flavor and $CP$ violation significantly in excess of Standard Model predictions. Strong bounds arising from $K$-$\overline K$ mixing particularly constrain additional flavor violation in the first two generations.

\item {\it Naturalness of the Higgs mass in light of collider bounds.}

 The Higgs mass should lie above observational bounds without requiring significant tuning of the scale of electroweak symmetry breaking (EWSB). Though supersymmetry may solve the problem of quadratically divergent contributions to the Higgs mass, the remaining logarithmic contributions are often problematic.

\item {\it Freedom from coincidence problems (technically natural or otherwise).}

EWSB in supersymmetric generalizations of the Standard Model 
requires both supersymmetric and SUSY-breaking mass parameters. The scale of ostensibly supersymmetric contributions  (e.g., the $\mu$-term in the MSSM) should naturally coincide with that of non-supersymmetric contributions.
 \end{enumerate}
There are also a variety of other curious features of the Standard Model that might ideally be explained by a supersymmetric extension. While such desiderata are numerous, perhaps the three most attractive possibilities would be:
\begin{itemize}
\item Improved SM predictions for perturbative gauge coupling unification.
\item A plausible dark matter candidate.
\item An explanation of the fermionic flavor hierarchy and the CKM matrix (the SM flavor puzzle).
\end{itemize}
Though these are by no means necessary requirements of a supersymmetric extension of the Standard Model, their natural realization would be quite compelling. Perturbative gauge coupling unification and a reasonable dark matter candidate tend to arise in many supersymmetric models, but explanations for the Standard Model flavor puzzle are more elusive.

It is common in model building to place a particular emphasis on precision constraints, as they directly rule out large classes of models.  This focus naturally favors gauge mediation~\cite{Dine:1981gu,Dimopoulos:1981au,AlvarezGaume:1981wy,Nappi:1982hm,Dine:1982zb,Dine:1994vc,Dine:1995ag} as it readily explains the smallness of new flavor-violating contributions to Standard Model processes via the flavor-blindness of gauge interactions.  While gauge mediation beautifully explains precision data, it often has difficulties accommodating other required or desirable features of supersymmetric models.  For example, the simplest realizations of gauge mediation predict a spectrum of soft masses with significant splittings between colored and electroweak scalars; soft spectra consistent with collider bounds typically lead to heavy colored states whose logarithmic contributions to the Higgs mass erode the naturalness of electroweak symmetry breaking. Even more troublesome is the coincidence between the $\mu$-term and the soft masses that is required for electroweak symmetry breaking. The $\mu$-term appears in the superpotential
\beq
W = \mu H_u H_d \ ,
\eeq   
and is therefore perfectly supersymmetric and unrelated to the scale of soft masses. It seems to be a striking coincidence of the MSSM that two {\it a priori} unrelated scales are of the same order of magnitude.  The simplest attempts to produce the $\mu$-term dynamically frequently lead to an unacceptably large $B\mu$-term~\cite{Dvali:1996cu}. A number of solutions have been proposed to circumvent this problem (e.g.~\cite{Hall:2002up, Giudice:2007ca,Roy:2007nz,Murayama:2007ge,Csaki:2008sr,Komargodski:2008ax,Green:2009mx,Dine:2009swa,Evans:2010ru,SchaferNameki:2010mg}), but they typically require a variety of additional 
assumptions.  These shortcomings suggest that gauge mediation, if a feature of Nature, may not be realized in its most conventional form.

In this paper, we will attempt to broaden the context of supersymmetry breaking so as to retain many of the attractive features of gauge mediation while providing explanations for the origin of the flavor hierarchy and natural electroweak symmetry breaking.  Fundamentally, both the flavor puzzle and the $\mu$-problem require explaining the size of superpotential couplings; this suggests a common origin for the two phenomena. In each case, one must start by explaining the absence of large tree level couplings.

 We will show that a particular example of deconstruction~\cite{ArkaniHamed:2001ca,Hill:2000mu} (more specifically deconstructed gaugino mediation) provides a natural framework for extending conventional gauge mediation to address the flavor puzzle and the $\mu$-problem.  In deconstructed models, the gauge group at high energies is enlarged to two, or more, SM-like gauge groups.  At low energies, the groups are spontaneously broken to the diagonal SM subgroup by the VEVs of bifundamental link fields, $\chi_i$ and $\tilde \chi_i$.  The extended gauge symmetry of the model can provide a natural explanation for the absence of superpotential couplings.  We will use precisely this mechanism to explain the hierarchy of Standard Model Yukawas and the small size of $\mu$. 

\begin{figure}[t]
\begin{center}
\epsfxsize=0.9\textwidth\epsfbox{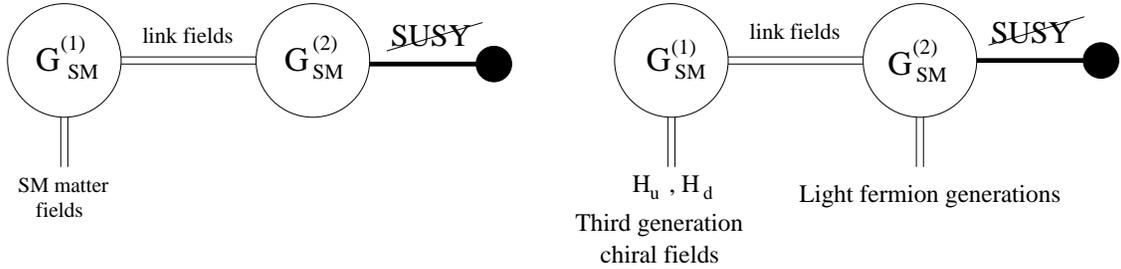}
\caption{Pure gaugino mediated scenario (illustration on the left) vs. the two-site model of flavor described in section \ref{section:modelC} (illustration on the right).}
\label{fig:modelC}
\end{center}
\end{figure}

Deconstruction is a compelling framework for building models of supersymmetry breaking and mediation.  Specifically, deconstruction provides a mechanism for suppressing the scalar masses generated by gauge mediation (i.e., gaugino mediation).  Although the original models of gaugino mediation~\cite{Kaplan:1999ac,Chacko:1999mi} employed spatial separation in an extra dimension to suppress scalar masses, the same effect was achieved in four dimensions (as with so many extra-dimensional phenomena) via a deconstruction~\cite{Csaki:2001em,Cheng:2001an}.  In the simplest version of deconstructed gaugino mediation the gauge group in the UV is extended to $G^{(1)}_{SM} \times G^{(2)}_{SM}$. The messengers of the SUSY-breaking sector are charged under $G^{(2)}_{SM}$, but not under $G^{(1)}_{SM}$, while the SM matter fields are all charged under $G^{(1)}_{SM}$ (see illustration on the left hand side of Fig~\ref{fig:modelC}).  If the link fields get VEVs $\langle \chi \rangle $ that are parametrically smaller than the messenger mass scale $M$, the soft masses of the MSSM scalars are suppressed by $\frac{\langle \chi \rangle^2 }{M^2}$.  The gaugino masses are unaffected by the VEVs and are the same as in gauge mediation.  It may seem that a model of this sort is fairly cumbersome and would be hard to naturally realize in a fully dynamical model.  To the contrary, the necessary collection of gauge groups, link fields, and symmetry-breaking potential were shown to arise dynamically in~\cite{Green:2010ww}; a simple deformation of massive SQCD leads to spontaneously broken SUSY and an IR realization of deconstructed gaugino mediation.

It is a small step from deconstructed gaugino mediation to a natural and flavorful supersymmetric Standard Model.  A simple way to introduce flavor into this setup is to move the light generations such that they are charged under the gauge group $G^{(2)}_{SM}$. 
Then the light renormalizable Yukawa terms (as well as the mixing terms between different generations) are simply not allowed by extended gauge invariance.\footnote{This possibility was first raised in the context of doublet-triplet splitting in~\cite{Witten:2001bf}.} In order to introduce these terms, operators of dimension 4 and higher should be added to the superpotential, suppressed by some flavor-physics scale $M_*$. After the link fields get VEVs, we find that the light Yukawas are formed, suppressed by powers of $\frac{\langle \chi \rangle }{M_*}$. This rearrangement has a significant impact on the resulting soft masses, as scalars charged under $G^{(2)}_{SM}$  receive masses from gauge mediation and are therefore far heavier than the  scalars  charged under $G^{(1)}_{SM}$. 

Of course, the idea of explaining the flavor puzzle within a dynamical model of SUSY-breaking is not new. Even before the construction of concrete models,  the crucial features of the supersymmetric spectrum were found to be intimately tied to Standard Model flavor~\cite{Dimopoulos:1995mi, Cohen:1996vb}. These ideas were explicitly realized in models of direct gauge mediation, where supersymmetry breaking and Standard Model flavor arise from the same dynamics~\cite{ArkaniHamed:1997fq,Luty:1998vr}.\footnote{For recent attempts to realize these ideas in the context of meta-stable SUSY-breaking in supersymmetric QCD~\cite{Intriligator:2006dd} see~\cite{Franco:2009wf,Craig:2009hf,Behbahani:2010wh}. An interesting attempt to solve the flavor puzzle and the $\mu$-problem simultaneously along these lines was undertaken in~\cite{SchaferNameki:2010iz}.} In this paper, we will pursue an alternate approach, leveraging the philosophy of deconstruction to build weakly-coupled, calculable models of flavorful supersymmetry breaking. 
 The same ingredients that govern the mediation of supersymmetry breaking also explain the flavor hierarchy, solve the $\mu$-problem, and render electroweak symmetry breaking natural. 
Our realization of flavor does not depend on the UV dynamics of the theory, so there is no obstacle to directly embedding our model in the construction of~\cite{Green:2010ww} or any other model that leads to the deconstruction ansatz in the IR.  

Although exploiting the two-site extension of the SM is attractive, we should be careful to point out an important limitation. With only two gauge groups, we will be unable to produce two different hierarchies between the quark masses; producing all hierarchies dynamically would require at least a three-site model. Therefore, one small number (usually the hierarchy between the first and the second generation quarks) essentially arises ``by hand''.\footnote{ A mass hierarchy may also arise from accidental degeneracies in the Yukawa texture. While these options may sound unappealing, one should remember that extending these models to three sites is straightforward and we mostly concentrate on the two-site models for simplicity. Of course, a three-site model is more strongly constrained by flavor-changing neutral currents. This is reminiscent  of challenges faced by flavor models in flat extra dimensions \cite{Hall:2001rz}.}  Similarly, one might be concerned that we are limited to explaining the smallness of at most two of three mixing angles in the CKM matrix. That said, the Cabibbo angle is fairly large, so perhaps this is all that Nature asks of us. 

As anticipated above, solving the $\mu$-problem is also natural in this framework. There are two possibilities, depending on the arrangement of Higgses between the two gauge groups. First, one can take a relatively conservative approach, assuming that both Higgses ($H_u$ and $H_d$) are charged under the gauge group $G^{(1)}_{SM}$, but the tree-level $\mu$-term is forbidden by some global symmetry.\footnote{One can also assume that the $\mu$-term is formed accidentally small at the high scale. Then its smallness is protected by a non-renormalization theorem.} Then the leading $\mu$-term arises by the insertion of link-fields in the superpotential. The resulting scale of $\mu$ is natural in a theory with a low scale of supersymmetry breaking. Having assigned the Higgses to $G^{(1)}_{SM}$, it is straightforward to determine the ideal distribution of the remaining SM chiral fields, guided by the observed flavor hierarchy and the fact that each SM generation is separately anomaly-free. This structure is realized in the ``vector Higgs'' model which we analyze in detail in section~\ref{section:modelC}.

The second approach to the $\mu$-problem is more novel: we assume that $H_u$ and $H_d$ are  \emph{charged under different gauge groups} and the $\mu$-term itself is forbidden by gauge invariance. This model very elegantly solves the $\mu$-problem of the MSSM, and addresses the flavor puzzle in highly non-standard way. Naively, the Yukawa matrices in this model predict an inverted hierarchy in the down-type quark sector and hence an untenable CKM matrix. However, due to a hierarchy in the Higgs sector, $m^2_{H_d} \gg | m_{H_u}^2| \gg B\mu$, we naturally obtain $\tan \beta \gg 1$. In this unusual regime of SUSY parameter space the down-type quark masses can be dominated by the coupling to $H_u^\dagger$ induced at one loop, rather than tree-level couplings to $H_d$~\cite{Dobrescu:2010mk}. Therefore the quark masses and mixing angles are sensitive to the hierarchy between the superpartners. The radiative couplings arising from the natural soft spectrum lead to viable fermion mass matrices and a realistic CKM matrix. This structure is characteristic of the ``chiral Higgs'' model  which we analyze in detail in section~\ref{section:modelD}.

The phenomenology of both these models is very exciting and quite different from the conventional MSSM paradigm. Indeed, the whole infrared structure of these supersymmetric Standard Models -- from Yukawa couplings to soft masses to electroweak parameters -- only emerges just above the TeV scale when the link fields acquire VEVs. As alluded to above, the scalars  charged under $G_{SM}^{(2)}$ get the usual gauge-mediated soft masses, while the masses of scalars charged under $G_{SM}^{(1)}$ are screened. 
This translates into the spectrum of ``more minimal SUSY''~\cite{Cohen:1996vb}, with the third generation scalars being relatively light, while the scalars of the first two generations are significantly heavier. In the chiral model this feature has an important impact on fermion masses, as we have already emphasized.  Both models alleviate the fine-tuning of conventional gauge and gaugino mediation while still preserving a form of gauge coupling unification. The low scale of Higgsing in these theories limits the size of contributions to the Higgs mass coming from stop loops. Moreover, significant nonsupersymmetric corrections to the $SU(2)_L \times U(1)_Y$ $D$-terms naturally lift the Higgs mass above observational limits without reliance on radiative corrections.

Note that neither model we present here is flavor blind.\footnote{Unlike deconstructed gaugino mediation, our models do not fall into the class of general gauge mediation, as defined in~\cite{Meade:2008wd}. Therefore we should not expect them to automatically share generic features of gauge mediation, even though supersymmetry breaking is mediated by gauge interactions.}  However, these models are triply-protected against  excessive flavor violation. First, the mixing angles in the scalar masses are closely related to the CKM-matrix mixing angles and therefore are naturally small. Second, the first two generations of soft masses are identical, ameliorating the most stringent bounds from flavor-changing neutral currents (FCNCs). Third, any remnant FCNCs involving the light generations are further suppressed since the scalar superpartners of these generations are very heavy. This security against prohibitive flavor violation arises precisely because flavor and supersymmetry breaking are so closely intertwined.

Our paper is organized as follows. In section~\ref{sec:direct} we review general facts about two-site models, as well as important features that arise in dynamical models of SUSY-breaking. In section~\ref{section:modelC} we discuss the model with vector-like Higgses and analyze its important predictions. In section~\ref{section:modelD} we turn to the model with chiral Higgses. Finally in section~\ref{sec:conclusions} we conclude. Some technical details and numerical results regarding FCNCs, soft spectra, and unification are relegated to appendices.   
\section{The Standard Model via moose}\label{sec:direct}

Before we turn to the concrete models outlined in the introduction, let us begin by discussing a few general features of two-site quivers leading to the Standard Model in the infrared. Such two-site quivers offer the simplest realization of deconstruction, and suffice to capture all of the features relevant to our models.  The ultraviolet theory consists of two gauge groups, $G^{(1)}_{SM} \times G^{(2)}_{SM}$, which we will take to each consist of $SU(3)_i \times SU(2)_i \times U(1)_i$.  The fields that appear in the Standard Model will be charged under either $G^{(1)}_{SM}$ or $G^{(2)}_{SM}$ in their usual representations. In addition to the Standard Model fields, the theory contains bifundamental link fields $\chi, \tilde \chi$ with a superpotential of the form
\beq\label{eq:sb}
W_\chi = A(\chi \tilde \chi - f^2) \ ,
\eeq
where $f$ is a constant with dimensions of mass and $A$ is essentially a Lagrange multiplier field; this superpotential may naturally arise from confining dynamics~\cite{Green:2010ww}. 

The soft spectrum of Standard Model fields depends sensitively on how supersymmetry breaking is communicated to the two sites. We will focus on the case where supersymmetry breaking is communicated via messengers of gauge mediation charged under $G^{(2)}_{SM}$; this leads  to the usual soft masses for $G^{(2)}_{SM}$ matter, including positive-definite soft masses $m_{\chi, \tilde \chi}^2$ for the link fields.  The scalar potential for the link fields then depends on both supersymmetric and nonsupersymmetric contributions. When $f^2 > m_{\chi, \tilde \chi}^2$, the superpotential (\ref{eq:sb})  leads to VEVs $\langle \chi \rangle, \langle \tilde \chi \rangle$ such that $G^{(1)}_{SM} \times G^{(2)}_{SM} \to \left (SU(3)_c \times SU(2)_L \times U(1)_Y \right)_{\rm diagonal}$, identified with the Standard Model gauge symmetries in the infrared. The Standard Model gauge couplings are determined by matching at the scale of higgsing, 
\beq
\frac{1}{g_i^2 } = \frac{1}{g_{(1)i}^2} + \frac{1}{g_{(2)i }^2} \ .
\eeq
where $i \in 1,2,3$ labels the gauge group $U(1)$, $SU(2)$ or $SU(3)$.  The UV couplings $g_{(1)i}$ and $g_{(2)i}$ are not necessarily equal at the scale of higgsing.  It will therefore be convenient to define the angles
\beq
\sin \theta_i = \frac{g_{(1)i}}{\sqrt{g_{(1)i}^2 + g_{(2)i}^2}} \hspace{1cm} \cos \theta_i = \frac{g_{(2)i}}{\sqrt{g_{(1)i}^2 + g_{(2)i}^2}}
\eeq
such that $g_{i} = g_{(1)i} \cos \theta_i = g_{(2)i} \sin \theta_i$, where the $\theta_i$ are not necessarily identical.

The groups $G_{SM}^{(i)}$ may or may not unify into a full $SU(5)_i$ at some higher scale; if only one group unifies, some of the successful features of orbifold GUTs may be realized \cite{Csaki:2001qm}. In the models we will consider, it is possible to unify at least one of the groups in the UV, thereby preserving a sense of gauge coupling unification; we reserve a more detailed discussion for Appendix A. 

\subsection{Energy Scales and Soft Masses}

Although we will present the resulting soft spectra in full detail in subsequent sections, let us sketch the salient shared features here. The gauginos of $G^{(2)}_{SM}$ obtain masses at one loop via gauge mediation. The gauginos of $G^{(1)}_{SM}$ then obtain masses from tree-level mixing with those of $G^{(2)}_{SM}$ at the scale of higgsing. In the limit $m_\lambda \ll f$, the light gaugino mass eigenstates in the infrared are the superpartners of SM gauge bosons with masses $m_{\lambda_i} \simeq \sin^2 \theta_i \; m_{\lambda_{(2)i}}$. Consequently, if the messengers of SUSY-breaking have a universal mass ($M$), then the SM gauginos will obey unified mass relations. The size of the gaugino masses relative to the remaining soft masses depends on whether gaugino masses are generated at leading order -- $m_{\lambda} \propto F/M$ -- or subleading order -- $m_{\lambda} \propto F^3/M^5$ -- in SUSY-breaking, with SUSY-breaking scale $F^{1/2}$. This is determined by the precise details of supersymmetry breaking and has significant influence on the low-energy spectrum.  

If the gaugino masses arise at leading order in SUSY-breaking, we have the usual result $m_{\lambda_i} = \frac{\alpha_i}{4\pi} \frac{F}{M}$.  In principle, the SUSY-breaking scale, $F$, can be quite large in this case, although we require the link field VEV $\langle \chi \rangle \sim 10-50$ TeV in order to explain the size of the $\mu$-term.  The scalar masses for matter charged under $G_{SM}^{(2)}$ are of the same order as the gaugino masses, $m^2_{(2)} \sim m_{\lambda}^2$, although they may be logarithmically enhanced by RG flow.  The scalar masses for matter charged under $G_{SM}^{(1)}$ are generated at 3 loops and are therefore lighter than the gauginos, $m^2_{(1)} \sim (\frac{1}{10} m_{\lambda})^2$.  The 3-loop contributions depend in part on the soft masses $m_{\chi, \tilde \chi}^2$, which will prove significant for the sign of $m_{H_u}^2$.  RG flow does not significantly enhance the soft masses $m_{(1)}^2$ because of the low scale of higgsing.  A more detailed analysis of the soft masses in deconstructed gaugino mediation can be found in \cite{McGarrie:2010qr, Sudano:2010vt, Auzzi:2010mb,Auzzi:2010xc,McGarrie:2011dc}.

If instead the gaugino masses arise at subleading order in SUSY-breaking, we have $m_{\lambda_i} \sim 0.1 \frac{\alpha_i}{4\pi} \frac{F^3}{M^5}$. In this case the SUSY-breaking scale is necessarily low in order to avoid fine-tuning and therefore $F \sim M^2 \sim (10^3~{\rm TeV})^2$.  The scalar masses for matter charged under $G_{SM}^{(2)}$ are now much larger than the gaugino masses, $m^2_{(2)} \sim (10\, m_{\lambda})^2$.\footnote{Consequently, the heavy scalars lie around $3-10$ TeV, depending on the suppression of gaugino masses. Intriguingly, heavy squarks in the $3-5$ TeV range may be discoverable with $\leq$ 100 fb$^{-1}$ at the LHC \cite{Fan:2011jc}.}  The scalar masses for matter charged under $G_{SM}^{(1)}$ are generated at 3 loops and are of the same order as the gaugino masses, $m^2_{(1)} \sim m_{\lambda}^2$. 

We will consider both possibilities in our models.  In the first model (the vector-like Higgs), both possibilities are acceptable and only differ in the detailed spectra.  Our second model (the chiral Higgs) will require that the gaugino masses are {\it suppressed} in order to yield a reasonable flavor hierarchy. Because SUSY-breaking occurs at a low scale, the LSP is the gravitino in both cases, with mass in the cosmologically safe eV - keV range.

Having outlined the general features of two-site models, let us devote a moment to discussing their ultraviolet origin.

\subsection{On the origin of mooses}

As infrared effective theories, deconstructed gaugino mediation and related quiver-based models are complete. Yet their intricate structure of gauge groups, link fields, higgsing, and supersymmetry breaking may seem somewhat {\it ad hoc} from an ultraviolet perspective. However, it is possible for the architecture of deconstruction and supersymmetry breaking to arise {\it dynamically} at low energies from a strongly-interacting gauge theory~\cite{Green:2010ww}. This has the appealing feature of explaining the myriad features of deconstructed models in terms of fairly generic gauge dynamics. 

Though not necessary to realize the models described below, such dynamical origins do lead to a number of useful and distinctive features that are not automatically realized by a purely IR effective theory.  While it is somewhat outside the scope of this paper to review the complete setup of dynamically deconstructed theories (we refer the reader to \cite{Green:2010ww} for details), it is worth briefly mentioning some qualitative features characteristic of a dynamic scenario:

\begin{itemize}
\item The masses of $G_{SM}^{(2)}$ gauginos tend to come out significantly smaller than the corresponding soft masses of scalars charged under $G_{SM}^{(2)}$. This happens due to the IR stability of the vacuum, leading to a parametric suppression of gaugino masses $m_{\lambda} \propto F^3/M^5$~\cite{Komargodski:2009jf}.\footnote{This problem can potentially be circumvented if one utilizes IR-metastable vacua (see e.g.~\cite{Kitano:2006xg,Giveon:2009yu,Koschade:2009qu,Barnard:2009ir,Maru:2010yx,Curtin:2010ku}), however these models tend to be quite complicated. Note also that this problem has nothing to do with the R-symmetry and one finds suppressed gaugino masses even if R-symmetry is maximally broken as in~\cite{Giveon:2008ne}. }
\item {\it A priori} SUSY-breaking can occur at any scale. However, if one relies on the IR-stable vacuum, SUSY-breaking and its mediation are required to occur at the lowest allowed scale, as the masses $M$ of messengers charged under $G_{SM}^{(2)}$ are of the same order as $\sqrt{F}$, the scale of primordial SUSY-breaking.
\item Certain small dimensionless numbers arise naturally. In deconstructed theories where $H_u$ and $H_d$ are charged under different gauge groups, the $\mu$-term arises when the link field $\chi$ acquires a VEV; since $\langle \chi \rangle \sim 10-50$ TeV, the operator generating the $\mu$-term often requires an $\mathcal{O}(0.1-0.01)$ coefficient to be plausible. However, in dynamical models, the coupling $\chi H_u H_d$ originates as a dimension-four operator involving a field in the SUSY-breaking sector. As a result, a small coefficient arises naturally via an effective operator of the form $\delta W \sim \frac{\sqrt{F}}{M_*} \chi H_u H_d,$ where $\sqrt{F}/M_* \sim 0.1-0.01$ leads to the necessary suppression.

\end{itemize}

Throughout the remainder of the paper, we will remain agnostic about whether or not the effective theories under consideration arise dynamically. 


\section{Vector-like Higgses} \label{section:modelC}

In this section we introduce the first model that simultaneously addresses the SM flavor puzzle and solves the $\mu/B\mu$ problem. Here we adopt a conservative approach to both problems: we will leave the Higgs fields charged under the same gauge group.

To realize a model of flavor, different generations are given different charges under $G_{SM}^{(1)}$ and $G_{SM}^{(2)}$.  For a given set of charges, only certain marginal Yukawa couplings will be gauge invariant.  Given that the flavor hierarchies are similar in the up and down sector, it is reasonable to charge $H_u$ and $H_d$ under the same group.  This configuration is schematically illustrated -- and contrasted with conventional deconstructed gaugino mediation -- in Fig~\ref{fig:modelC}. From the picture alone it is easy to understand intuitively the origin of the flavor hierarchies. One can write down marginal Yukawa couplings between the Higgses and the heavy flavors, since all these fields are charged under $G_{SM}^{(1)}$. However,
it is impossible to introduce similar terms for the light flavors by virtue of gauge invariance, since the light flavors are charged under $G_{SM}^{(2)}$. In order to form these terms, one should insert powers of the link fields. Consequently, these Yukawas will be suppressed by $\left( \frac{\chi}{M_*} \right)^n$, where the power $n$ depends on details of the setup and $M_*$ is some high scale whose nature we will discuss in more detail below. 

Our approach to the $\mu$-problem is fairly conventional:  we will forbid a tree level $\mu$-term via, e.g., a PQ symmetry, and instead produce $\mu$ from the VEV of some dynamical field. In our model we have a natural candidate for this dynamical field: the link fields. Indeed, the $\mu$-term may be formed by the operator
\beq\label{mu:C}
W \sim \frac{\chi \tilde \chi H_u H_d}{M_*}~. 
\eeq 
This leads to an effective $\mu$-term,  $\mu_{\rm eff} \sim \frac{f^2}{M_*}$. Although one might be concerned about generating a dangerous PQ axion when $\chi, \tilde \chi$ get VEVs, the form of the full link field superpotential provides a sufficiently large source of explicit PQ breaking to render the would-be PQ axion harmless.
 
 One can easily see that this mild modification of the deconstruction ansatz has the following generic predictions:
\begin{itemize}
\item quark masses are hierarchical, 
\item angles in the CKM matrix are small,
\item $\tan \beta$ is moderate,
\item lighter fermions have heavier sfermionic partners.
\end{itemize}   
Let us now render these predictions in more complete detail.

\subsection{Layout}
Consider two copies of the SM gauge groups, $G_{SM}^{(1)}$ and $G_{SM}^{(2)}$. The messengers of SUSY-breaking are charged under the gauge group $G_{SM}^{(2)}$ and the link fields $\chi,\, \tilde \chi$ are in bifundamental representations of the SM gauge groups. We charge the Higgses and the third generation chiral fields under the gauge group $G_{SM}^{(1)}$, precisely as in deconstructed gaugino mediation. However the first and the second generation fields are charged under $G_{SM}^{(2)}$.\footnote{The quark mass matrices are not diagonal in this basis and therefore strictly speaking we cannot refer to these fields as well-defined generations. However, since we also predict small quark mixing angles, this is a fairly good first approximation.} A full field alignment is given in table~\ref{tab:modelC}.   
\begin{table}[t]
\centering 
\begin{tabular}{|c||c|c|} \hline
 Chiral field & $SU(3)_1\times SU(2)_1\times U(1)_1$ &  $SU(3)_2\times SU(2)_2\times U(1)_2$ \\ \hline
$\chi_h$ & $(3,1,-\frac{1}{3})$ & $ (\bar 3,1,\frac{1}{3})$ \\
$\tilde \chi_h$ & $ (\bar 3,1,\frac{1}{3})$ &  $(3,1,-\frac{1}{3})$ \\
$\chi_l$ & $(1,2, \frac{1}{2})$ & $(1, 2, -\frac{1}{2})$ \\
$\tilde \chi_l$ & $(1, 2, -\frac{1}{2})$ & $(1,2, \frac{1}{2})$ \\
$H_u$ & $(1, 2, \frac{1}{2})$ & singlet \\
$H_d$ & $(1,2, -\frac{1}{2})$ & singlet \\
$Q_3$ & $ (3,2, \frac{1}{6})$ & singlet \\
$Q_{1,2}$ & singlet & $(3,2,\frac{1}{6})$ \\
$\bar u_3$ & $(\bar 3, 1, -\frac{2}{3})$ & singlet \\
$\bar u_{1,2}$ & singlet & $(\bar 3, 1, -\frac{2}{3})$ \\
$\bar d_3$ & $(\bar 3, 1, \frac{1}{3})$ & singlet \\
$\bar d_{1,2}$ & singlet & $(\bar 3, 1, \frac{1}{3})$\\
$L_3$ & $(1, 2, -\frac{1}{2})$ & singlet \\
$L_{1,2}$ & singlet &  $(1, 2, -\frac{1}{2})$ \\
$\bar e_3$ & $ (1,1,1)$ & singlet \\
$\bar e_{1,2}$ & singlet & $(1,1,1)$ \\ \hline
\end{tabular} 
\caption{Alignment of chiral fields in the two-site model with vector-like Higgses.}
\label{tab:modelC}
\end{table} 

First, let us discuss the flavor structure in this model.\footnote{Throughout the paper, we will focus on flavor in the quark sector, since the measured values of CKM matrix elements provide the strongest constraint on flavor textures; extending our analysis to the lepton sector is straightforward.} At the level of renormalizable operators only the third generation Yukawas are allowed. In order to have masses for the light generations, as well as mixing terms between different generations, we must involve the link fields. For example, the second generation mass is formed by introducing the following operators in the superpotential:
\beq\label{flavormix:C}
\Delta W \sim \frac{H_u \tilde \chi_l Q_2 \bar u_2}{M_*} + \frac{H_d \chi_l Q_2 \bar d_2 }{M_*}~.
\eeq 
Similar operators should be introduced for the first generation fermions and the mixing terms. For further convenience we define two small parameters:
\beq
\epsilon_l \equiv \frac{\langle \chi_l \rangle }{M_*} = \frac{\langle \tilde \chi_l \rangle}{M_*}, \ \ \ \ 
\epsilon_h \equiv \frac{\langle \chi_h \rangle }{M_*} = \frac{\langle \tilde \chi_h \rangle}{M_*}~. 
\eeq  
In terms of these small parameters one finds the following Yukawa textures in a  two-site vector-like Higgs model:
\beq
Y_u \sim \sin \beta \left( 
\begin{array}{ccc}
 \epsilon_l & \epsilon_l & \epsilon_h \epsilon_l\\
 \epsilon_l & \epsilon_l & \epsilon_h \epsilon_l \\
 \epsilon_h^2 & \epsilon_h^2 & 1
\end{array} \right), \ \ \ \
Y_d \sim \cos \beta \left(
\begin{array}{ccc}
 \epsilon_l & \epsilon_l & \epsilon_h \epsilon_l \\
 \epsilon_l & \epsilon_l & \epsilon_h \epsilon_l \\
 \epsilon_h & \epsilon_h & 1 \\
\end{array} \right) \ . 
\eeq 
Note that hereafter the matrix entries merely represent the parametric suppression, rather than numerical behavior; there should naturally be $\mathcal{O}(1)$ numerical coefficients in each entry. However, there is no need for a significant hierarchy in these coefficients; the flavor hierarchy is dictated entirely by gauge invariance.

As we have explained in the introduction, any two-site model is incapable of producing two different hierarchies in the masses and three small angles. This model is not an exception: it produces a hierarchy between the third generation quarks and the light generations, and it explains two small angles in the CKM matrix. In particular, our alignment predicts:
\beqa 
m_{u,c} \propto \sin \beta\ \epsilon_l v, &  \ \ \ &  
m_t \propto \sin \beta\ v \\
m_{d,s} \propto \cos \beta\ \epsilon_l v, & \ \ \ & 
m_b \propto \cos \beta\ v~.   
\eeqa   
The CKM matrix elements $V_{td},V_{ts},V_{ub},V_{cb}$ are small. More specifically, we predict the following parametric structure of the CKM matrix:
\beq\label{CKM:modelC}
V_{\rm CKM} \sim \left( \begin{array}{ccc}
                     1 & 1 & \epsilon_h \epsilon_l \\
                     1 & 1 & \epsilon_h \epsilon_l \\
                     \epsilon_h \epsilon_l & \epsilon_h \epsilon_l & 1
                    \end{array} 
\right) \ .
\eeq
By construction it cannot explain the hierarchy between the light generations or the  smallness of the Cabbibo angle -- though ultimately $\sin \theta_c \sim 0.23$ is not terribly small. In practice, this construction just loosely reproduces the observed quark mass hierarchies and mixing angles; some numbers should be adjusted ``by hand'', perhaps through near-degeneracies in the $\CO(1)$ coefficients. One can partially resolve some of these problems by extending this model to be a three-site model along the lines we present here. Here we continue with the two-site model for the sake of simplicity, which perfectly illustrates the relevant phenomenological features and avoids the most stringent bounds from FCNCs.

\subsection{Phenomenology and soft spectrum}
Matching the Yukawa textures to observed masses and mixings dictates the hierarchy of scales in this model. First, the order of magnitude of $\epsilon_l$ is determined from the observed ratios of quark masses:
\beq
\epsilon_l \sim \frac{m_c}{m_t} \sim \frac{m_s}{m_b}~. 
\eeq 
This ratio is of order $\CO (10^{-2})$. On the other hand, we should choose $\epsilon_h $ such that the  CKM  matrix is reproduced closely enough. Then it is natural to choose $\epsilon_h \sim \CO (10^{-1})$. Of course these are just very crude estimations and altering $\mathcal{O}(1)$ numbers allows for $\epsilon_h$ and $\epsilon_l$ to be the same order of magnitude. Given that it is natural that $\epsilon_l \sim \epsilon_h$, we will further imagine 
\beq
\epsilon_l \sim \epsilon_h \lesssim \CO (10^{-2}) \ ,
\eeq 
and rely on various numerical factors to adjust $V_{ub}$ and $V_{cb}$ to the observed values. 

As we have already explained in section~\ref{sec:direct}, it is at least conceivable that these models involve low-scale SUSY-breaking. In this particular model we have strong reasons for the SUSY-breaking scale to be low. If we assume that operators in~\eqref{flavormix:C} responsible for the flavor structure are produced at the same scale as the effective $\mu$-term~\eqref{mu:C}, we find that $\mu_{\rm eff} \sim \epsilon \langle \chi \rangle $.  Under this assumption, we require the following hierarchy of scales:
\beq
\langle \chi_l \rangle \sim 10 - 20\ {\rm TeV}, \ \ \ \ M_* \lesssim 10^3\ {\rm TeV}~. 
\eeq     
The necessary mass scale $M_*$ is quite suggestive of the messenger mass in low-scale mediation. Indeed, it is easy to envision various scenarios in which the $\mu$-term and flavor interactions are generated by additional couplings of SM fields to messenger-scale matter, or even to messengers themselves. This is particularly easy to realize in dynamical models, where many of the fields responsible for the deconstruction ansatz lie around the messenger scale and may be appropriately coupled to Standard Model and link fields. Generating the couplings in (\ref{mu:C}) and (\ref{flavormix:C}) by integrating out messenger-scale matter provides added rationale for low-scale SUSY-breaking in the 100 - 1000 TeV range.

Let us now discuss the superpartner spectrum. Unlike in models of gauge and gaugino mediation, the scalar mass matrices are not proportional to the identity matrix, and in the fermion mass eigenbasis they are not diagonal. However, these matrices are \emph{exactly diagonal} in the flavor basis. The scalars of the first two generations interact directly with the messengers via regular two-loop diagrams of gauge-mediation and obtain identical masses:
\beq
m_{GM}^2 \sim \left( \frac{\alpha}{4\pi} \right)^2 \left( \frac{F}{M}\right)^2~.
\eeq 
In contrast, the mass-squared of the third generation scalars are further suppressed by a factor $\langle \chi \rangle^2 / M^2$ since it effectively comes from gaugino rather than gauge mediation:
\beq
m_{\tilde g M}^2 \sim \left( \frac{\alpha}{4\pi} \right)^2 \left( \frac{\langle \chi \rangle}{M} \right)^2 \left( \frac{F}{M}\right)^2~.
\eeq 
 There is also a three-loop contribution to this soft mass that can exceed, in some cases, the two-loop $\langle \chi \rangle $-suppressed contribution~\cite{DeSimone:2008gm}. Since the squark masses for the first two generations are proportional to the identity matrix, the only sizable mixing in the soft mass matrix is between the third generation squarks and the light generations. The explicit form of the LH squark mass-squared matrix is
\beq
m^2_{\tilde Q} \sim \left( \begin{array}{ccc}
   m_{GM}^2 & 0 & \epsilon_h \epsilon_l m_{GM}^2 \\
   0 & m_{GM}^2 & \epsilon_h \epsilon_l m_{GM}^2 \\
   \epsilon_l \epsilon_h m_{GM}^2 & \epsilon_l \epsilon_h m_{GM}^2 & m_{\tilde g M}^2                       
                        \end{array}\right) \ ,
\eeq         
while for the RH squarks we obtain somewhat larger mixings:
\beq
m^2_{\tilde {\bar u}} \sim  \left( \begin{array}{ccc}
   m_{GM}^2 & 0 & \epsilon_h^2 m_{GM}^2 \\
   0 & m_{GM}^2 & \epsilon_h^2  m_{GM}^2 \\
   \epsilon_h^2 m_{GM}^2 & \epsilon_h^2 m_{GM}^2 & m_{\tilde g M}^2                       
                        \end{array}\right), \ \ \ 
m^2_{\tilde {\bar d}} \sim  \left( \begin{array}{ccc}
   m_{GM}^2 & 0 & \epsilon_h m_{GM}^2 \\
   0 & m_{GM}^2 & \epsilon_h  m_{GM}^2 \\
   \epsilon_h m_{GM}^2 & \epsilon_h m_{GM}^2 & m_{\tilde g M}^2                       
                        \end{array}\right) \ .
\eeq   
Since the $G_{SM}^{(1)}$ soft masses are screened and do not run significantly, it is relatively easy to obtain a numerical spectrum in which collider bounds are satisfied while keeping third-generation squarks below 1 TeV.

Despite the smallness of the $G_{SM}^{(1)}$ soft masses and limited RG evolution between the Higgsing scale and the weak scale, $m_{H_u}^2$ is negative over a wide range of parameters to ensure electroweak symmetry breaking ~\cite{DeSimone:2008gm}. In particular, the negative three-loop contribution to the Higgs soft mass beats the positive two-loop contribution due to the size of $\lambda_t$ and $g_3$.

As described in section \ref{sec:direct}, the gaugino masses may arise at leading or sub-leading order in the SUSY-breaking scale ($F$).  Neither possibility is preferred in this model, as the scalar mass spectrum is not significantly altered by the gauginos due to the short RG flow.  Of course, the detailed phenomenology and signatures will depend on the size of the gaugino masses.  

Finally, let us mention an amusing correction to the Higgs quartic coupling that arises in this model. As it has been noticed in~\cite{Maloney:2004rc, DeSimone:2008gm}, the D-terms of heavy $Z'$s and $W'$s also contribute to the Higgs quartic coupling. Since this contribution would  vanish if the breaking $G_{SM}^{(1)} \times G_{SM}^{(2)} \to G_{SM}$ were supersymmetric, it must be proportional to the soft masses of the link fields $\chi, \tilde \chi$:
\beq\label{Higgs4}
\delta V = \frac{g^2 \Delta}{8} \left| H_u^\dagger \sigma^a H_u + H_d^\dagger \sigma^a H_d \right|^2 + \frac{3 {g'}^2 \Delta'}{40}
\left| H_u^\dagger H_u - H_d^\dagger H_d\right|^2
\eeq      
where
\beq
\Delta = \frac{g_1^2}{g_2^2} \frac{2\tilde m_{\chi}^2}{M_2^2 + 2\tilde m_{\chi}^2} \ .
\eeq 
A similar formula pertains for $\Delta'$; here $M_2$ stands for the heavy $SU(2)$ gauge boson mass. Note that this way to ``fix'' the Higgs quartic coupling is extremely safe, since the D-terms preserve the custodial symmetry and therefore we do not find any corrections to the $T$-parameter. 
The corrections~\eqref{Higgs4} can be $\mathcal{O}(1)$ and mitigate the little hierarchy problem.\footnote{In the concrete examples of low scale gaugino mediation~\cite{DeSimone:2008gm} this correction was found to be negligibly small and did not have a strong impact on the Higgs mass bound. However, it was assumed in~\cite{DeSimone:2008gm} that the gauge group $G_{SM}^{(2)}$ is a full $SU(5)$ and therefore all the gauge couplings of that group unify; we are typically interested in relaxing this constraint.  Moreover, the ratio $g_1/g_2$ in conventional gaugino mediation is expected to be small in order to avoid an early Landau pole for $G_{SM}^{(1)}$. But here, having split SM fields between the two gauge groups, we find more matter charged under $G_{SM}^{(2)}$ rather than $G_{SM}^{(1)}$ and therefore we may easily have $\frac{g_1}{g_2} \gtrsim 1$. } The mass of the Higgs at tree level is bounded in this case by
\beq
m_h^2 \leq m_Z^2 + \frac{g^2 \Delta + {g'}^2 \Delta'}{2} v^2 \ .
\eeq   
This bound is almost saturated in the large $\tan \beta$ limit. As a result we can expect stops to be significantly lighter than 1 TeV, since the indirect bounds on their masses from the Higgs mass do not apply.  Thus both forms of the little hierarchy problem are solved -- the Higgs mass is naturally raised above collider bounds, while small logarithmic radiative corrections minimize tuning.


\subsection{Constraints}
As with any supersymmetric theory that gives rise to non-universal soft masses and new degrees of freedom near the TeV scale, we must determine what soft spectra may remain consistent with constraints coming from FCNCs and other considerations. There are three potentially significant contributions to FCNCs in the theories we are considering:

\begin{enumerate}

\item Conventional one-loop MSSM diagrams arising from the misalignment of the fermion and sfermion mass matrices, of which the gluino-mediated diagram is dominant.

\item Additional box diagrams involving the link fields $\chi, \tilde \chi$.

\item Tree-level FCNCs induced by the heavy gauge bosons of $G^{(1)}_{SM} \times G^{(2)}_{SM}$.

\end{enumerate}

Ultimately, the dominant contribution in the theories at hand arises from typical gluino-mediated box diagrams, proportional to generation-mixing soft masses. The contributions from box diagrams involving link fields are suppressed by $(v/M_*)^4$ and thus insignificant. The tree-level diagrams involving heavy gauge bosons are suppressed by the relevant elements of $V_{\rm CKM}$, and are readily compatible with experimental limits when the heavy gauge boson masses are all $M_i \gtrsim 10$ TeV (for a general discussion of tree-level FCNCs mediated by heavy gauge bosons, see \cite{Delgado:1999sv}; for recent limits on the relevant effective operators see, e.g., \cite{Bona:2007vi}). These scales also ensure that contributions to the $\rho$ parameter from triplet link fields (and violations of custodial symmetry by both the link fields and heavy gauge bosons) are well within current limits~\cite{DeSimone:2008gm}.

The most stringent constraints on our model will arise from the flavor- and CP-violating contributions to FCNCs from the misaligned sfermion masses.  Before we delve into the details, let us first outline the most important results. We find that CP-conserving limits from $K^0 - \overline{K}^0$ and $B^0 - \overline{B}^0$ mixing are almost equally important and put very mild constraints on our models that are readily satisfied. On the other hand, if there is a maximal complex phase in the rotation matrices for the right-handed quarks, constraints from the CP-violating parameter $\epsilon_K$ are satisfied by only a narrow range of soft masses. We must therefore either assume some phase alignment or restrict ourselves to the allowed range of soft masses. All other constraints from FCNC processes pose weaker limits.


We may parametrize the gluino-mediated contributions to FCNCs using the mass insertion method, in which limits may be placed directly on the normalized splittings $(\delta_{ij}^{\tilde f})\equiv \sum_k (U_{\tilde f})_{ik} (U_{\tilde f})_{kj}^* \approx m_{\tilde f,ij}^2/\langle  m^2_{\tilde f} \rangle$, where, e.g., $U_{\tilde f}$ is the unitary matrix diagonalizing the {\it sfermion} mass matrix 
 in the {\it fermion} mass eigenbasis. To compute numerical limits we use expressions for the gluino-mediated contributions to meson mixing appropriate to a hierarchical soft spectrum, which we obtain from \cite{Bertolini:1990if} using the techniques of \cite{Giudice:2008uk}.\footnote{We consider only chirality-preserving insertions $\delta^{\tilde f}_{LL, RR}$; chirality-mixing insertions $\delta^{\tilde f}_{LR, RL}$ arise from $A$-terms that are loop-suppressed in the models under consideration.} This gives contributions to meson mixing at the scale of the soft masses; limits are computed by RG evolving these contributions to the relevant IR scale using the expressions in \cite{Bona:2007vi}.
The primary constraints on our spectrum arise in principle from $K^0 - \overline{K}^0$, $D^0 - \overline{D}^0$, and $B^0 - \overline{B}^0$ mixing 
as well as the rare decays $\mu\to e\gamma$ and $b \to s \gamma.$

The Standard Model contributions to meson mixings fall within the measured values, but depend on hadronic uncertainties to an extent that the full contribution is unknown; thus we may take as our constraints the requirement that new physics contributions to $\Delta m_K, \Delta m_D, \Delta m_B,$ $\Gamma(b \to s \gamma)$, and $\Gamma(\mu \to e \gamma)$  do not exceed (in magnitude) the measured values, which we take from~\cite{Nakamura:2010zzi} (excepting $\Gamma(b \to s \gamma)$, which we infer from the $BR(B \to X_s \gamma)$ value in~\cite{TheHeavyFlavorAveragingGroup:2010qj}). We may also compute limits on the CP violating parameter $\epsilon_K$, assuming the presence of an additional phase in the right-handed mixing angles. Here the prediction for the Standard Model contribution is fairly robust, so we may limit the new physics contribution by requiring that it be smaller than the difference between the observed value of $|\epsilon_K|$ (also from~\cite{Nakamura:2010zzi}) and a Standard Model contribution $|\epsilon_K|_{SM} \approx 1.80 \times 10^{-3}$~\cite{Buras:2008nn}.  The limits are listed in Table~\ref{tab:fcnclimits}.

\begin{table}[t]
\begin{center}
\begin{tabular}{|c|c|} \hline
Process & Limit  \\ \hline
$\Delta m_K$ & $(3.483 \pm 0.006) \times 10^{-15}$ GeV  \\
$\Delta m_D$ & $(1.57^{+.39}_{-.41} ) \times 10^{-14}$ GeV  \\
$\Delta m_B$ & $(3.337 \pm 0.033) \times 10^{-13}$ GeV  \\
$ BR(B \to X_s \gamma)$ & $(355 \pm 24 \pm 9) \times 10^{-6}$  \\
$BR(\mu \to e\gamma)$ & $< 1.2 \times 10^{-11}$  \\
$|\epsilon_K|$ & $2.228 \times 10^{-3}$ \\
\hline
\end{tabular}
\end{center}
\caption{Relevant FCNC processes and limits}
\label{tab:fcnclimits}
\end{table}

Parametrically, the SUSY contribution to the mixing of a meson $M$ comprised of valence quarks $q_i, q_j$ scales as 
\beq
\Delta m_M \approx \alpha_3^2 m_M f_M^2 \frac{(\delta^q_{ij})^2}{\tilde m^2} \ ,
\eeq
provided that $m_{\tilde g} \approx \tilde m$.
Similarly
\begin{eqnarray}
BR(b \to s \gamma) &\approx& \alpha_3^2 \alpha \frac{m_b^4}{\tilde m^4}  (\delta^d_{23})^2 \times m_b \tau_B \ , \\
BR(\mu \to e \gamma) &\approx& \alpha^3 \frac{m_W^4}{\tilde m^4} (\delta^l_{12})^2 \times BR(\mu \to e \nu_\mu \nu_e).
\end{eqnarray}

For the specific vector Higgs model presented above, constraints arising from FCNCs are milder than one would naively expect; the soft masses for the first two generations of squarks and sleptons are universal at the high scale, and differences arise only due to negligible RG flow to the weak scale. At leading order, $ (\delta^d_{12})_{LL,RR} \approx (\delta^u_{12})_{LL,RR} \approx (\delta^l_{12})_{LL,RR} \approx 0$. Consequently, contributions to $\Delta m_D$, and $\Gamma(\mu \to e \gamma)$ are negligible. The contribution to $\Delta m_K$ comes at the {\it next to leading order} in $\delta$ (with four insertions, $(\delta^d_{13} \delta^d_{23})^2$). This dominates over the leading order both because the mixings $\delta^d_{13}, \delta^d_{23}$ significantly exceed $\delta^d_{12}$, and because the LO diagram is suppressed relative to the NLO diagram by a factor $m_{\tilde b}^2/m_{\tilde d}^2$ from the heavy sfermion propagator. In principle, the NLO contribution still provides a significant bound, since the limits on $\Delta m_K$ and $\epsilon_K$ are so stringent.\footnote{We thank Gilad Perez for bringing this to our attention.} Indeed, one may readily estimate the contribution and find it is in the ballpark of the current experimental bound. Other relevant constraints come from the limits $\Delta m_B$ and $\Gamma(b \to s \gamma)$. Parametrically, $(\delta_{13}^d)_{LL}, (\delta_{23}^d)_{LL} \propto \epsilon_l \epsilon_h$ and $(\delta_{13}^d)_{RR}, (\delta_{23}^d)_{RR} \propto \epsilon_h$ are sufficient for both bounds to be readily satisfied for a wide range of soft masses.  A detailed numerical illustration of the allowed parameter space is shown in Appendix B; there we show explicitly that the leading order constraints on $\Delta m_B$ and the next-to-leading order constraints on $\Delta m_K$ are almost equally important, and both are satisfied over the full range of soft masses (for our choice of $\epsilon_h,\ \epsilon_l$ and order-one numbers in Yukawa matrices, though these limits are comparable for different choices of numerical parameters). Stronger limits from $\epsilon_K$ arise in the presence of an additional $\mathcal{O}(1)$ complex phase unaligned with the phase of the CKM matrix. We find that there is still a range of soft masses that satisfy these limits for $\mathcal{O}(1)$ complex phase, but outside this range some degree of alignment is required between phases.

By contrast, in a three-site variation of this model, no symmetry relates the soft masses for any of the three generations, leading to much more stringent constraints from $\Delta m_K$~and~$\Delta m_D$.

Such are the limits arising from FCNCs; there are also considerations of naturalness and positivity of soft masses. The mass of the stop is, of course, bounded from above by the naturalness of electroweak symmetry breaking; this leads us to favor stop masses (and, by extension, the soft masses of all colored scalars charged under $G_{SM}^{(1)}$) $\lesssim 1$ TeV. Due to the smallness of the first- and second-generation Yukawa couplings, there is no such constraint on, e.g. the charm and up squark masses. There is, however, an upper bound on the masses of first- and second-generation squarks due to their (negative) two-loop influence on the stop mass renormalization. Though this is screened by the VEVs of the link fields, it is nonetheless prudent to limit first- and second-generation squark masses to $\lesssim 20$ TeV \cite{ArkaniHamed:1997ab}.

\section{Chiral Higgs model}\label{section:modelD}

The model presented in the previous section is very appealing from a flavor point of view, and offers a solution to the $\mu$-problem through the VEVs of the link fields.  The main difference between the vector-like Higgs model and conventional deconstructed gaugino mediation is that the Standard Model fields are split between the different sites.  A model of flavor arises naturally because gauge invariance forbids certain renormalizable Yukawa couplings.  

Explaining the flavor hierarchy and the small size of the $\mu$-term are fairly similar problems, in the sense that both require an origin for small, technically natural superpotential parameters.  It is therefore not surprising that the origin of flavor discussed in the previous section suggests a similar solution to the $\mu$-problem.  If we again use a two-site model of gaugino mediation, there is a natural solution to the $\mu$-problem where the Higgs fields are split between the two sites such that gauge invariance forbids a tree level $\mu$-term.  One instead includes the renormalizable superpotential 
\beq
W \sim \chi_l H_u H_d \ ,
\eeq
such that $\mu \sim \langle \chi \rangle$ after the two gauge groups are broken to the diagonal. This solution is attractive in that it offers a new explanation for the absence of a tree-level supersymmetric $\mu$-term (gauge invariance) and makes use of existing ingredients already present in the theory. Of course, the scale $\langle \chi \rangle$ is not {\it a priori} the same as the weak scale, and in this setup requires a numerical coefficient on the order of $0.1-0.01$ to be completely natural.\footnote{In this sense it is completely analogous to $\mu$ in 5D gaugino mediation, for which a similar numerical coefficient is necessary \cite{Schmaltz:2000gy}.}  However, as discussed earlier, in dynamical models such as~\cite{Green:2010ww} both the scale of $\langle \chi \rangle$ and the numeric coefficient are natural because of the required low scale of SUSY-breaking and origin of the necessary operators.

An interesting consequence of moving $H_d$ is that it introduces a large hierarchy in the Higgs soft masses.  Specifically, $m^2_{H_d}$ is generated through gauge mediation while $m^2_{H_u}$ is generated by gaugino mediation.  Furthermore, $B \mu$ is generated only by MSSM RG flow and is therefore smaller than $\mu^2$.  As a result, electroweak symmetry breaking necessarily occurs at $\tan \beta \sim 10^2 - 10^4$ and naturally yields an uplifted Higgs scenario.  We will be most interested in $\tan \beta \sim 10^4$ which arises naturally when the gaugino masses are suppressed.

The extreme size of $\tan \beta$ implies that the masses of the bottom quark and tau lepton are generated predominantly by loop-induced, non-holomorphic couplings to $H_u^{\dag}$.  As a result, the texture of the Yukawa coupling to $H_d$ does not completely determine the flavor hierarchies and CKM matrix of the Standard Model.  The hierarchy of soft masses in the down sector also contributes significantly to the flavor structure through loop-generated couplings to $H_u^{\dag}$; in this sense, fermionic flavor is induced at one loop by sfermionic flavor.

\subsection{Phenomenology and soft spectrum}
Let us now turn to the parametric behavior of masses and mixing angles that arise in the chiral Higgs model.  For the sake of clarity we will reserve a more detailed discussion of certain features for Appendix C.

We will consider a two-site deconstruction model with two copies of the SM gauge groups, $G_{SM}^{(1)}$ and $G_{SM}^{(2)}$, and a vector-like pair of bifundamental link fields $\chi$ and $\tilde{\chi}$.  As in the previous section, we will only couple SUSY-breaking to $G_{SM}^{(2)}$ via gauge mediation and therefore soft masses for fields charged under $G_{SM}^{(1)}$ will be suppressed.  To achieve a light stop we should charge the top under  $G_{SM}^{(1)}$.  We should similarly charge $H_u$ under $SU(2)_{(1)}$ in order to allow for a tree level top Yukawa.  We would like to forbid a $\mu$-term for the Higgses by gauge invariance and therefore $H_d$ will be charged under $SU(2)_{(2)}$.

With the Higgs fields split between the two nodes, anomaly cancellation will not allow us to split the down quarks and leptons in the same way.  Specifically, one can no longer organize the fields into $SU(5)$ multiplets to determine the anomaly free combinations.  Of course, one can still find anomaly free configurations.  Indeed, one anomaly free possibility is to split the fields as in Table~\ref{tab:modelD}, by moving one lepton doublet to compensate for moving $H_d$.\footnote{One could always add additional matter to cancel anomalies. However, in the spirit of minimality, we find it more interesting to pursue the configurations allowed by SM chiral matter alone.}

\begin{table}[t]
\centering 
\begin{tabular}{|c||c|c|} \hline
 Chiral field & $SU(3)_1\times SU(2)_1\times U(1)_1$ &  $SU(3)_2\times SU(2)_2\times U(1)_2$ \\ \hline
$\chi_h$ & $(3,1,-\frac{1}{3})$ & $ (\bar 3,1,\frac{1}{3})$ \\
$\tilde \chi_h$ & $ (\bar 3,1,\frac{1}{3})$ &  $(3,1,-\frac{1}{3})$ \\
$\chi_l$ & $(1,2, \frac{1}{2})$ & $(1, 2, -\frac{1}{2})$ \\
$\tilde \chi_l$ & $(1, 2, -\frac{1}{2})$ & $(1,2, \frac{1}{2})$ \\
$H_u$ & $(1, 2, \frac{1}{2})$ & singlet \\
$H_d$ & singlet & $(1,2, -\frac{1}{2})$ \\
$Q_3$ & $ (3,2, \frac{1}{6})$ & singlet \\
$Q_{1,2}$ & singlet & $(3,2,\frac{1}{6})$ \\
$\bar u_3$ & $(\bar 3, 1, -\frac{2}{3})$ & singlet \\
$\bar u_{1,2}$ & singlet & $(\bar 3, 1, -\frac{2}{3})$ \\
$\bar d_{3}$ & $(\bar 3, 1, \frac{1}{3})$ & singlet \\
$\bar d_{1,2}$ & singlet & $(\bar 3, 1, \frac{1}{3})$\\
$L_{3,2}$ & $(1, 2, -\frac{1}{2})$ & singlet \\
$L_{1}$ & singlet &  $(1, 2, -\frac{1}{2})$ \\
$\bar e_{3}$ & $ (1,1,1)$ & singlet \\
$\bar e_{1,2}$ & singlet & $(1,1,1)$ \\ \hline
\end{tabular} 
\caption{Alignment of the Chiral fields in the two-site model with chiral Higgses.}
\label{tab:modelD}
\end{table}

The primary motivation for splitting the Higgses is to address the $\mu$-problem.  Because the $\mu$-term is generated by the operator $\chi_l H_u H_d$, the scale of Higgsing is necessarily the TeV scale and the $B \mu$-term is not generated at leading order.  The down-type Higgs is charged under $G_{SM}^{(2)}$ and receives a gauge mediated mass $m^2_{H_{d}} \sim \frac{\alpha^2}{(4 \pi)^2} \frac{F^2}{M^2}$.  The up-type Higgs is charged only under $G_{SM}^{(1)}$ and its soft mass is suppressed.  As before, our gauge mediated soft masses will be at least 1 - 10 TeV to accomodate collider bounds on the lighter fields.  The resulting Higgs sector soft masses obey the hierarchy
\beq
\label{eqn:higgshier}
m_{H_d}^2 \gg  \mu^2 \sim | m_{H_u}^2| \gg B\mu \ .  
\eeq
Because of the large hierarchy between the masses, we are necessarily considering a model with extremely large values of $\tan \beta$.

One can make a reasonable estimate of $\tan \beta$ from the RG equations for $B \mu$.  At the scale $\langle \chi_l\rangle$, $B \mu = 0$ because there is no F-term for $\chi_l$.  MSSM RG flow generates a non-zero $B\mu$-term proportional to $\mu$ and the wino and bino masses.  The RG flow is short so we can estimate of the size of $B\mu$ by
\beq
\label{eqn:Bmu}
B \mu \sim - \mu\left( \frac{3  \alpha_2}{2 \pi} m_{\tilde W} \log \frac{\chi_l}{m_{\tilde W}} + \frac{\alpha}{2 \pi} m_{\tilde B}\log \frac{\chi_l}{m_{\tilde B}} \right) \ .
\eeq
The value of $\chi_l \sim 10-20$ TeV is fixed in order to explain the small size of the $\mu$-term.  As a result, there is not sufficient RG time to generate a large value of $B \mu$, and instead we find $B \mu \sim 0.1 \mu m_{\lambda}$. When the Higgs soft masses obey the hierarchy in (\ref{eqn:higgshier}), we can write $B \mu$ and $\mu^2$ as
\beq
\label{eqn:ewsymmb}
B\mu \tan \beta \simeq m_{H_d}^2  \qquad {\rm and} \qquad \mu^2 = -m_{H_u}^2 - \frac{M_Z^2}{2} \ . 
\eeq
Electroweak symmetry breaking requires $m_{H_u}^2 < 0$ which can be achieved by a combination of  the three-loop soft terms and RG flow.  As a result, the $\mu$-term is required to be small for most choices of parameters.

\begin{figure}[t] 
   \centering
   \includegraphics[width=3in]{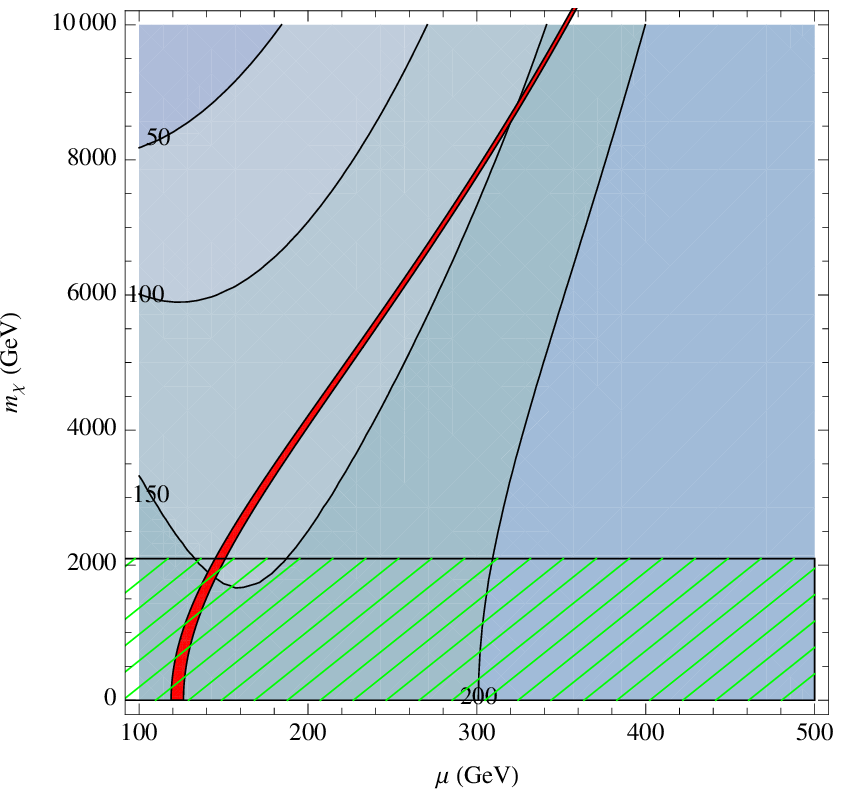} 
   \includegraphics[width=3in]{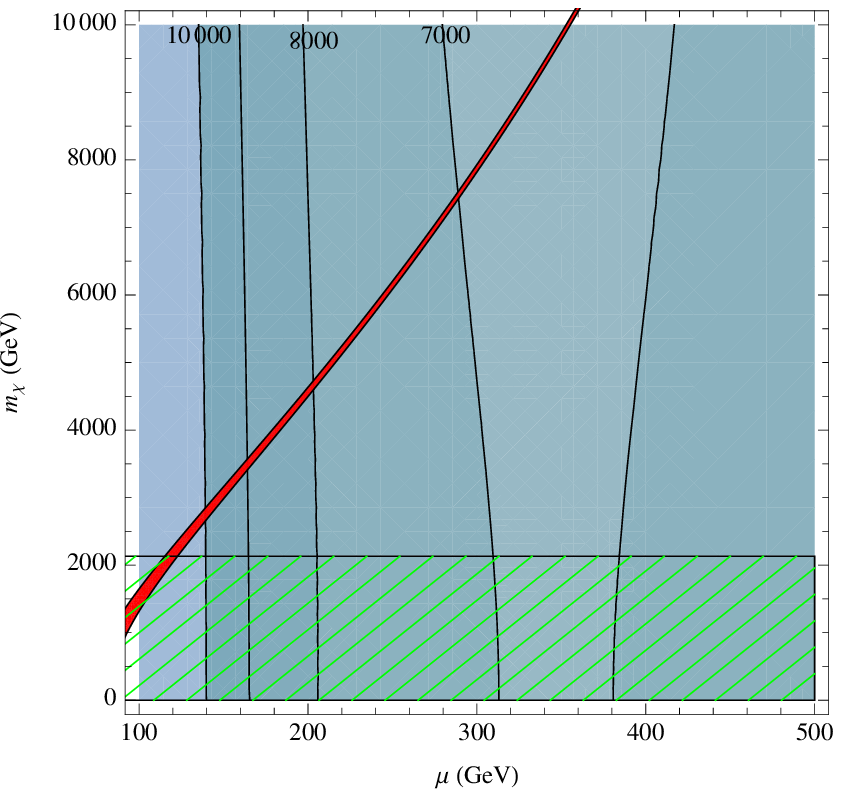} 
   \caption{(Left) Typical values of $\tan \beta$ as a function of $\mu$ and the link field soft mass, assuming no suppression of gaugino masses and $F/M = 100$ TeV with $\theta_1 = \frac{\pi}{4}, \theta_2 = \frac{\pi}{5}, \theta_3 = \frac{\pi}{5}$. The overlaid red stripe indicates parameters leading to the observed value of $m_Z$. The green striped region at low $\tilde m_{\chi}$ is excluded by LEP bounds on $\tilde \tau_R$ NLSP. Bounds on $\mu$ from chargino mass limits vary between $90-200$ GeV in this scenario \cite{Meade:2009qv}, and are not shown. (Right) Typical values of $\tan \beta$ assuming $\mathcal{O}(0.1)$ suppression of gaugino masses and $F/M = 700$ TeV with $\theta_1 = \frac{\pi}{4}, \theta_2 = \frac{\pi}{5}, \theta_3 = \frac{\pi}{5}$.}
   \label{fig:ewsb}
\end{figure}

The precise value of $\tan \beta$ depends sensitively on whether or not the gaugino masses are suppressed. Let us first assume that gaugino masses are generated at leading order in SUSY-breaking ($m_{\lambda} \propto F /M$).  In this case, there is no hierarchy between the gaugino masses and the down type soft mass (i.e. $m_{\tilde W}^2 \simeq m_{H_d}^2$).  $B\mu$ is numerically small because it is only generated by a short RG flow, as in (\ref{eqn:Bmu}).  It is further suppressed relative to $m_{H_d}^2$ by the small size of $\mu$, which satisfies $\mu^2 \sim - m_{H_u}^2 \ll m_{H_d}^2$.   From our above estimate, we then expect $\tan \beta \sim 100$.  The values of $\tan \beta$ that are realized for unsuppressed gauginos are also shown in figure \ref{fig:ewsb} and confirm this estimate.  
If instead the gaugino masses arise at subleading order in SUSY-breaking, then $m_{H_d}^2 \gtrsim (10 \, m_{\tilde W})^2$.  In this case, $B \mu$ is suppressed both by the small size of the gaugino masses and $\mu$.  Due to the limited RG flow, it is difficult to produce large negative values of $m_{H_u}^2$.  As a result, $\mu^2 \sim - m^2_{H_u} \ll m_{\lambda}^2$, even though gaugino masses are suppressed.  Using (\ref{eqn:ewsymmb}), we find
\beq
\tan \beta \sim \CO (10^4) \ .
\eeq
The range of $\tan \beta$ that is realized in this model is shown in figure \ref{fig:ewsb}.  As we will see below, these large values of $\tan \beta$ will be necessary for a viable model of fermionic flavor, and therefore we will assume the gaugino masses arise at subleading order in the chiral Higgs model.

It is reasonable to wonder how the bottom quark mass might be generated if $\tan \beta$ is so large. Typically, perturbativity of the bottom Yukawa coupling restricts electroweak symmetry breaking in the MSSM and related theories to $\tan \beta \lesssim 50$. However, such theories remain perfectly viable even when $\tan \beta \gg 50$. In this case the primary source for the mass of the bottom quark and the tau lepton is no longer tree-level couplings to $H_d$, but rather nonholomorphic couplings to $H_u^\dag$ induced at one loop after supersymmetry is broken. This region of electroweak symmetry breaking parameters has been dubbed the ``uplifted'' supersymmetric region \cite{Dobrescu:2010mk}, of which the chiral Higgs model presented here is a prototypical example. We will explain the salient features as they arise, but refer the reader to \cite{Dobrescu:2010mk} for more details. 

As in the previous section, we assume that all gauge invariant operators allowed in the superpotential arise with order one coefficients.   Because of the charge assignments given  in table~\ref{tab:modelD}, the couplings $Y_u H_{u} Q_i \bar u_j$ and $Y_d H_d Q_i \bar d_j$ are given by
\begin{eqnarray}
Y_u =  \left( \begin{array}{ccc}
\epsilon_l & \epsilon_l & \epsilon_h^2 \\
\epsilon_l & \epsilon_l & \epsilon_h^2 \\
\epsilon_h^2 & \epsilon_h^2 & 1 
\end{array} \right) \ ,  \hspace{1cm} 
Y_d =  \left( \begin{array}{ccc}
1 & 1 & \epsilon'_h \\
1 & 1 & \epsilon'_h \\
\epsilon'_l \epsilon'_h & \epsilon'_l \epsilon'_h & \epsilon'_l
\end{array} \right) \ , 
\end{eqnarray} 
where $\epsilon_{l,h}= \chi_{l,h} / M_*$ and $\epsilon'_{l,h}= \chi_{l,h} / M'_*$.  Our choice of splitting the up-type and down-type flavor scales, $M_*$ and $M'_*$, is technically natural and may be useful for achieving a reasonably large bottom mass.

In the uplifted region, $\cos\beta \simeq (\tan \beta)^{-1} \ll 1$ and therefore the couplings to $H_d$ do not necessarily give the largest contribution to the physical quark masses in the down sector.   After including SUSY-breaking, there are also one-loop contributions to the down sector masses through the effective operators
\beq
\mathcal{L} \supset - \hat Y_d \bar d H^\dag_u Q - \hat Y_l \bar e  H_u^\dag L + {\rm h.c.}
\eeq
Since these effective operators arise from nonsupersymmetric loops, they depend sensitively on the scalar soft masses. 

In the flavor basis, the soft masses for $Q_i$ and $\bar{d}_i$ arise from gauge / gaugino mediation and take the form
\beqa
m_{\tilde{Q}}^2 \sim 
\left( \begin{array}{ccc}
m_{GM}^2 & 0 & 0 \\
0 & m_{GM}^2 & 0\\
0 & 0& m_{\tilde gM}^2
\end{array} \right) \ ,
 \hspace{1cm} 
m_{\tilde{\bar{d}}}^2 \sim
\left( \begin{array}{ccc}
m_{GM}^2 & 0 & 0 \\
0 & m_{GM}^2 & 0\\
0 & 0& m_{\tilde g M}^2
\end{array} \right) .
\eeqa
By integrating out the squarks, one generates the coupling $\hat Y_d \bar d H^\dag_u Q $ with coefficients
\begin{eqnarray}
\hat{Y}_d \sim \frac{\alpha_3}{\pi} \left( \begin{array}{ccc}
\delta &\delta& \epsilon'_h \delta \\
 \delta & \delta & \epsilon'_h \delta  \\
 \epsilon'_l \epsilon'_h \delta  & \epsilon'_l \epsilon'_h\delta  &  \epsilon'_l 
\end{array} \right) \ , 
\end{eqnarray}
where $\delta = m_{\tilde g M}^2 / m_{GM}^2 \sim 10^{-4}$ and we have used $\mu \sim m_{\lambda} \sim m_{\tilde g M}$ for the parametric estimate.

The physical masses of the down quarks and leptons in this model arise from both\footnote{A proper effective field theorist would integrate out the heavy Higgs, $H_d$, and describe all couplings in terms of $H_{u}^{\dag}$.  In that language, we are comparing the tree-level coupling suppressed by $B\mu / m_{H_d}^2$ to the loop-suppressed coupling.} classical couplings to $H_d$ and loop-induced couplings to $H_u^{\dag}$.  For $10^3 \lesssim \tan \beta \lesssim 10^4$, although the classical contributions are suppressed by $\tan\beta^{-1}$, their contribution is still larger than couplings suppressed by $\delta$.  As a result, we can write the down-type quark mass matrix as
\begin{eqnarray}
\label{eqn:downy}
M_d \simeq v \left( \begin{array}{ccc}
\cos \beta & \cos \beta & \epsilon'_h  \cos \beta  \\
 \cos \beta  &  \cos \beta  & \epsilon'_h   \cos \beta  \\
\epsilon'_l \epsilon'_h  \cos \beta  & \epsilon'_l \epsilon'_h  \cos \beta  & \frac{\alpha_3}{\pi} \epsilon'_l
\end{array} \right) \ .
\end{eqnarray} 
The resulting CKM matrix takes the form
\begin{eqnarray}
V_{\rm CKM} = \left( \begin{array}{ccc}
1 &1 & \epsilon^2 \\
1 &1 & \epsilon^2 \\
\epsilon^2 & \epsilon^2 & 1 
\end{array} \right) \ ,
\end{eqnarray}
where we have assumed $\epsilon_h = \epsilon_l = \epsilon$.

One of the remarkable features of this model is that the nice properties of the CKM matrix are driven by the hierarchy of the soft masses.  If the soft masses had been flavor blind, the flavor texture would have been controlled by $Y_d$ alone and would generically give rise to large mixing angles in the CKM matrix.  Instead, a realistic CKM matrix is generated primarily because the sbottom mass is parametrically lighter than the other down squark masses.  

One generic feature of these models is that the bottom quark (and the tau, which we have not discussed) are typically too light.  Our bottom mass is given by 
\beq
\label{eqn:bottommass}
m_b  \sim v \frac{2 \alpha_3}{3 \pi} \frac{\mu}{m_{\tilde Q_3}} \epsilon'_l \ ,
\eeq
where we have restored potentially significant order one numbers.  In order to explain the bottom mass, one would require that $\epsilon'_l \sim 1$ because of the loop suppression.  In principle, the ratio $\mu / m_{\tilde Q_3}$ could improve the situation, although this does not happen for the small values of $\mu$ necessary for electroweak symmetry breaking.  Regardless, in a controlled effective theory, the natural value of $\epsilon'_{l}$ will typically be smaller than is required to explain the observed bottom mass.  Therefore the $b$-quark mass remains an open question in this kind of model. 

Much as in the vector-Higgs model, the $D$-terms of heavy $Z'$s and $W'$s give rise to nonsupersymmetric, nondecoupling contributions to the Higgs quartic coupling. The calculation is analogous to that for the vector Higgs case, and results in a contribution to the scalar potential of the form
\beq
 \delta V =  \frac{g^4}{8} \frac{2 \tilde m_\chi^2}{M^2 + 2 \tilde m_\chi^2} \left( \frac{H_u^\dagger H_u}{\cos^2 \theta_2} + \frac{H_d^\dagger H_d}{\sin^2 \theta_2} \right)^2~ +  \frac{9 g'^4}{200} \frac{2 \tilde m_\chi^2}{M^2 + 2 \tilde m_\chi^2} \left( \frac{H_u^\dagger H_u}{\cos^2 \theta_1} + \frac{H_d^\dagger H_d}{\sin^2 \theta_1} \right)^2~,
\eeq 
Unlike in the previous case, this correction does not have the same functional structure as the D-term of the SM, due to the separation of Higgs fields between the two sites. Nonetheless, it suffices to lift the mass of the lightest neutral Higgs and mitigate the little hierarchy problem. 

\subsection{Constraints}

Given that the prediction for the bottom mass in the simplest chiral Higgs model comes out somewhat low, it is less useful to make precise numerical estimates of the parameter space allowed by FCNCs. However, the chiral model shares many of the same features that render the vector model safe from flavor constraints. Contributions to the $\rho$ parameter and custodial symmetry violation are small provided $M_i \gtrsim 10$ TeV. A combination of alignment, light-generation universality, and decoupling mitigate the contributions to meson mixing. Although separating the lepton doublets $L_1, L_2$ raises the specter of large $\mu \to e \gamma$, the large soft mass for $\tilde L_1$ renders this harmless.

\section{Conclusion}\label{sec:conclusions}

In this paper, we have presented models of supersymmetry breaking and mediation that (1) explain the Standard Model flavor hierarchy, (2)  naturally satisfy flavor constraints, (3) generate a reasonable $\mu$-term with no $\mu / B \mu$-problem, (4) preserve gauge coupling unification, and (5) solve the little hierarchy problem. These models do not require extensive epicycles of new phenomena, but rather generalize the paradigm of deconstructed gaugino mediation by charging Standard Model fields under two gauge groups in the ultraviolet. Qualitatively different phenomenology arises depending on the distribution of Higgs doublets between the two gauge groups. If the Higgses are vector-like (charged under the same gauge group), the flavor hierarchy arises naturally from the Yukawa couplings allowed by gauge invariance. The soft spectrum is of the more-minimal type, and electroweak symmetry breaking occurs at moderate values of $\tan \beta$. If the Higgses are chiral (charged under different gauge groups), the flavor hierarchy arises from a combination of allowed Yukawa couplings and radiative effects. The soft spectrum is likewise of the more-minimal type, but electroweak symmetry breaking occurs at high values of $\tan \beta$, such that radiative couplings to $H_u^\dag$ contribute significantly to down-type quark masses. These models comfortably satisfy collider constraints without extensive fine-tuning of electroweak symmetry breaking.

There are a number of open issues that we have not addressed, for the sake of concision, in the models presented here.  Firstly, the bottom mass in the chiral Higgs model is generically too small to be realistic.  The numerous appealing features of the chiral model make understanding the bottom mass worthwhile.  Secondly, we have discussed neither the lepton flavor hierarchy nor the PMNS matrix in any detail.  Both models naturally predict hierarchies in the lepton masses, but require additional input to address the neutrino spectrum. Both of these issues may be addressed by straightforward extensions of the framework presented here, and would fully realize the potential for deconstruction to provide a natural and flavorful supersymmetric Standard Model.

\acknowledgments{We are grateful to Kaustubh Agashe, Nima Arkani-Hamed, Guido Festuccia, Patrick Fox, Howard Georgi, Zohar Komargodski, John Mason, Gilad Perez, David Poland, Lisa Randall, Matthew Schwartz, Raman Sundrum, and Lian-tao Wang for useful discussions.  We also thank Daniel Baumann, Guido Festuccia and Zohar Komargodski for comments on the manuscript.  NC is supported in part by the NSF under grant PHY-0907744 and acknowledges the hospitality of the Aspen Center for Physics while this work was being completed. The research of DG~is supported by the DOE under grant number DE-FG02-90ER40542 and the Martin A.~and Helen Chooljian Membership at the Institute for Advanced Study.  DG~thanks the Berkeley Center for Theoretical Physics, the Stanford Institute for Theoretical Physics and the Centro de Ciencias de Benasque Pedro Pascual for hospitality while this work was being completed.  AK is partially supported by NSF grant PHY-0801323}.  

\appendix

\section{Remarks on gauge coupling unification}

\begin{figure}[t] 
   \centering
   \includegraphics[width=4in]{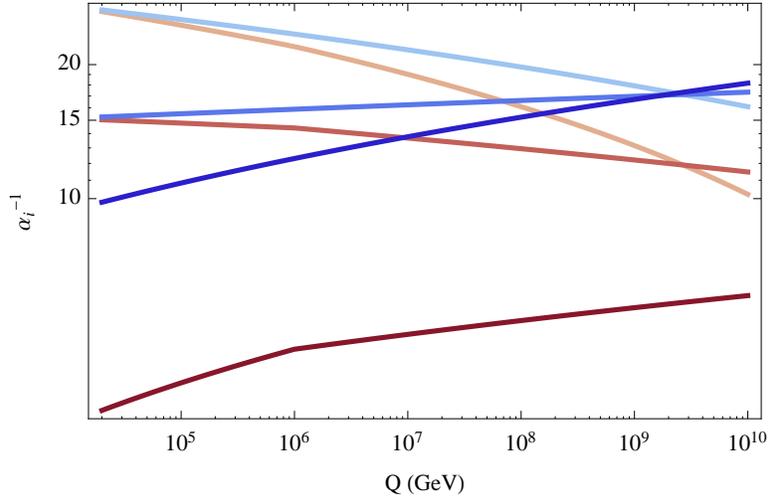} 
   \caption{One-loop running of $G_{SM}^{(1)}$ (blue) and $G_{SM}^{(2)}$ (red) inverse couplings $\alpha_i^{-1}(Q)$ for $\theta_1 = \theta_2 = \frac{\pi}{4}$, $\theta_3 = \frac{\pi}{6}$, $\langle \chi \rangle = 20$ TeV, and $\frac{\sqrt{F}}{M} = 10^3$ TeV, matched to measured SM gauge couplings at $\langle \chi \rangle$. Here $G_{SM}^{(1)}$ perturbatively unifies at $5 \times 10^9$ GeV.}
   \label{fig:uni}
\end{figure}

Gauge coupling unification of either $G_{SM}^{(1)}$ or $G_{SM}^{(2)}$ is fairly natural in both the vector and chiral Higgs scenarios (unification of both groups tends to require additional $SU(2)$-charged spectators). The unification of $G_{SM}^{(1)}$, in particular, is improved over the Standard Model prediction due to the additional matter charged under $SU(2)_{(1)}.$ In either case, the matching of $G_{SM}^{(1)} \times G_{SM}^{(2)}$ gauge couplings to the low-energy SM couplings -- plus the significant running contributions of link fields -- results in some form of accelerated unification \cite{ArkaniHamed:2001vr} around $5 \times 10^9$ GeV.  A typical scenario is illustrated in Fig.~\ref{fig:uni}.

Of course, such a low scale of unification raises usual questions about proton decay. Dimension-6 proton decay operators induced by $X,Y$ gauge bosons may be forbidden by unifying only $G_{SM}^{(1)} \to SU(5)_{(1)}$, as the light quarks are charged under $G_{SM}^{(2)}$ and then have no tree-level interactions with the resulting heavy gauge fields.  Triplet higgsino contributions to dimension-5 proton decay are somewhat more troublesome, as the appearance of such triplets at the apparent unification scale leads to a prohibitively high decay rate (and the requisite doublet-triplet splitting spoils the deconstructed solution to the $\mu$-problem). Avoiding these contributions requires that the Higgs fields do not fit into complete $SU(5)$ multiplets at the unification scale, reminiscent of familiar orbifold or stringy solutions. Alternately, such problems might be better resolved by trifinication, $G_{SM}^{(1)} \to [SU(3)]^3$. 

\section{Numerical flavor limits for the vector Higgs}

Here we will briefly present some numerical results for FCNCs in the vector-like Higgs model (the low prediction of $m_b$ in the chiral Higgs model makes a numerical rendering of parameter space less useful).  We take $\mathcal{O}(1)$ coefficients in the Yukawa matrix as follows:
\beq
Y_u \sim \sin \beta \left( 
\begin{array}{ccc}
 \epsilon_l & \epsilon_l & \epsilon_h \epsilon_l\\
 \epsilon_l & \epsilon_l & \epsilon_h \epsilon_l \\
 \epsilon_h^2 & \epsilon_h^2 & 1
\end{array} \right), \ \ \ \
Y_d \sim \cos \beta \left(
\begin{array}{ccc}
0.3 \epsilon_l & \epsilon_l & \epsilon_h \epsilon_l \\
 \epsilon_l & 1.3 \epsilon_l & \epsilon_h \epsilon_l \\
3 \epsilon_h & 4 \epsilon_h & 0.5 \\
\end{array} \right) \ .
\eeq 
With the choice $\epsilon_l = 0.02, \epsilon_h = 0.07$, this leads to a realistic CKM matrix:
\beq
V_{\rm CKM} \sim \left(
\begin{array}{ccc}
 -0.97 & 0.23& 0.009 \\
 0.23 & 0.97 & 0.035 \\
 0.0004 & -0.036 & 0.99
\end{array}
\right) \ .
\eeq
For $\tan \beta = 25$, this leads to good agreement with $m_t, m_b$; the predictions for $m_c, m_s$ are too large by a factor of a few, but this is not important for the determination of FCNC limits. As mentioned in the text, we compute numerical contributions to the one-loop gluino-mediated FCNC processes by expanding the one-loop amplitudes in \cite{Bertolini:1990if} for a hierarchical soft spectrum using the techniques of \cite{Giudice:2008uk}. We then RG evolve the resulting coefficients to the infrared using \cite{Bona:2007vi} and use them to bound the spectrum of soft masses.

\begin{figure}[htbp] 
   \centering
   \includegraphics[width=3in]{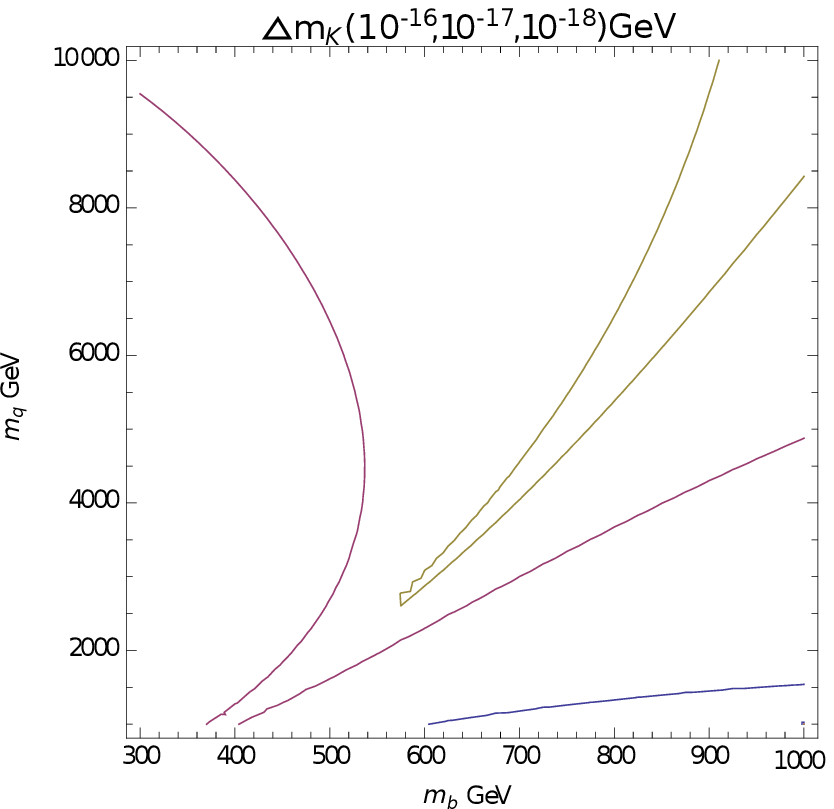} 
    \includegraphics[width=3in]{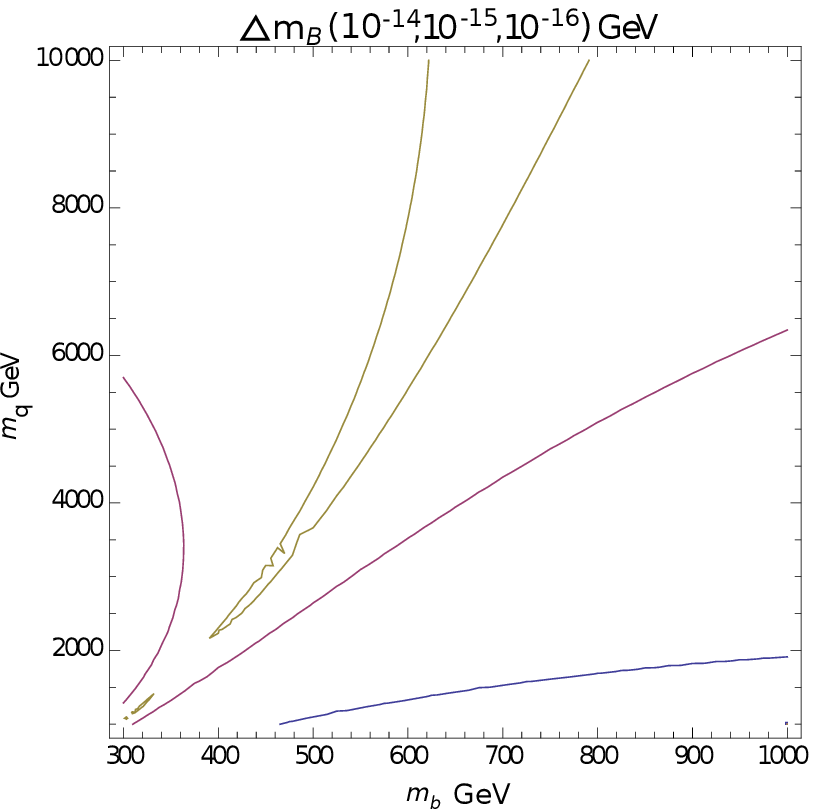} 
   \caption{Contours of $\Delta m_K$ (left) and $\Delta m_B$ (right) in GeV as a function of squark masses  {\it in the gauge eigenbasis} for the first two generations of down-type squark ($m_{\tilde q}$) and third generation ($m_{\tilde b}$), with $m_{\tilde g} = 800$ GeV. Limits on both mixings are satisfied over the full region of parameter space, assuming no additional CP violation.}
   \label{fig:kkbbmix}
\end{figure}

The primary limits come from the leading-order contribution to $B^0 - \overline{B}^0$ mixing and the next-to-leading-order contribution to $K^0 - \overline{K}^0$. Although the latter contributions are suppressed by an additional factor of $\delta^2$, the corresponding limits on $\Delta m_K$ are two orders of magnitude stronger than those on $\Delta m_B$, rendering the limits comparable. Ultimately, however, we find that both limits are readily satisfied by a wide range of parameters. Illustrative values of contributions to $\Delta m_K$ and $\Delta m_B$ are shown in figure \ref{fig:kkbbmix} as a function of the squark masses in the gauge eigenbasis, assuming no additional CP-violating phase. These contributions all lie below experimental bounds.

It is worth devoting a moment to considering both the small size of these contributions and their parametric dependence on the soft masses of the first two generations. In general the left-handed contributions to FCNCs (those proportional to $(\delta^d_{ij})_{LL}$) are highly suppressed by the smallness of the left-handed mixing angles and are negligible. The right-handed mixing angles are significantly larger, and by far the greatest contribution to FCNCs comes from terms proportional to $(\delta^d_{ij})_{RR}$. However, the largeness of these mixing angles also means that the right-handed sbottom mass in the fermion mass eigenbasis is increased by mixing with the heavier first two generations.\footnote{It is important to point out that this effect on third-generation soft masses is limited to $\tilde b_R$; in particular the masses of $\tilde b_L, \tilde t_L, \tilde t_R$ are essentially unchanged due to the smallness of the corresponding mixing angles.} This leads to a decoupling suppression of FCNCs as the masses of the scalars in the first two generations -- and hence also the mass of $\tilde b_R$ -- are raised. Consequently the right-handed contributions to FCNCs are suppressed by an unexpected form of decoupling. This also explains the mild dependence of $\Delta m_K$ and $\Delta m_B$ on the soft masses $m_{\tilde q}$, despite decoupling the scalars of the first two generations.

\begin{figure}[ht] 
  \centering
   \includegraphics[width=3in]{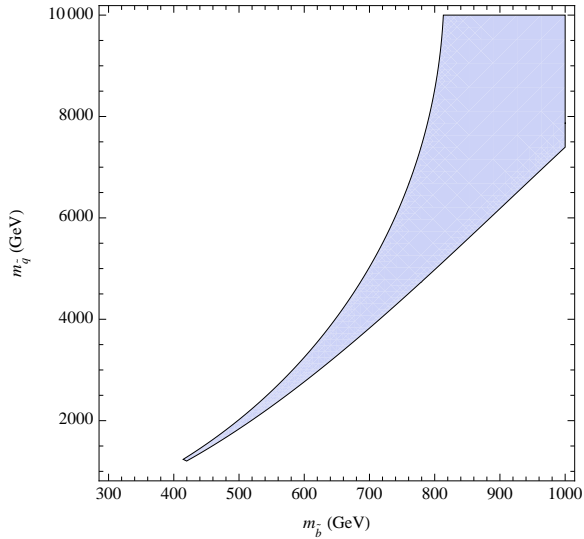} 
   \caption{Range of squark masses {\it in the gauge eigenbasis} allowed by limits on $\epsilon_K$ for $m_{\tilde g} = 800$ GeV, assuming an additional $\mathcal{O}(1)$ CP-violating phase.}
   \label{fig:cpv}
\end{figure}

More stringent constraints arise in the presence of additional CP violation due to the limits on $\epsilon_K$. {\it A priori}, there is nothing to prevent an additional CP violating phase in the right-handed mixing angles. We may compute the resulting constraints on soft masses from an $\mathcal{O}(1)$ phase unaligned with the phase in the CKM matrix, shown in figure \ref{fig:cpv}. Since the contribution to $\epsilon_K$ from Standard Model processes is fairly well known, we require the new contribution to be less than the difference between the observed value $|\epsilon_K| = 2.228 \times 10^{-3}$ and a representative Standard Model contribution, $|\epsilon_K|_{SM} \approx 1.80 \times 10^{-3}$. We find that there is still an allowed region of parameter space, thanks to the smallness of contributions to $\Delta m_K$. However, the full range of soft masses may only be realized if the new phase is aligned with the CKM phase to within $1-10 \%$.

\section{Numerical parameters for the chiral Higgs}

From our parametric estimates, we found that the chiral Higgs naturally is in a regime of $\tan \beta \gtrsim 1000$ and leads to a plausible flavor texture.  The details of the model are still sensitive to the precise value of $\tan \beta$ and the soft masses.  In this appendix we will explore the parameter space more quantitatively.  

The allowed values of $\tan \beta$ have a significant impact on the flavor structure in the down sector.  If $\tan \beta \sim 100$, then the $\cos \beta$ suppression in (\ref{eqn:downy}) is not sufficient to make $\bar d_3$ the heaviest quark.  The bottom quark is naturally associated with $\bar d_{1,2}$ which means there is no symmetry in the soft masses of the first two generations.  

The situation is much improved for $\tan \beta \sim 10^{3-4}$.  If we choose $\epsilon'_l \gtrsim 0.1$, the heaviest mass eigenstate is mostly aligned with $\bar d_3$.  This preserves that symmetry of the first two generations that is crucial for satisfying constraints from $K$-$\bar K$ mixing.  Explaining the observed hierarchy of down quark masses still requires further input in this case.  Specifically, if one wants to explain the hierarchy of the bottom to strange mass, one requires $\tan \beta \sim 10^4$.  Alternatively, one could also attribute the small bottom to strange ratio to the generically low value for the bottom mass, rather than insufficient suppression from $\tan \beta \sim 1000$.  

Typical values of $\tan \beta$ in the chiral Higgs model depend sensitively on the suppression of gaugino masses. If gaugino masses are unsuppressed relative to the scalars charged under $G_{SM}^{(2)}$, $\tan \beta \sim 70 - 200$ due primarily to the smallness of $B \mu$. However, if gaugino masses are suppressed, the hierarchy $m_{H_d}^2 \gg  \mu^2 \sim |m_{u}|^2 \gg B\mu$ leads to extreme uplifted regions with $\tan \beta \sim 10^4$. 

The size of $\tan \beta$ is closely related to the size of $\mu$, as should be clear from our parametric estimates.  Decreasing $\mu$ lowers the value of $B \mu$ and therefore increases $\tan \beta$.  Parametrically, larger $\tan \beta$ is desirable for explaining the bottom to strange ratio.  However, a more precise estimate of the bottom mass is given by (\ref{eqn:bottommass}), which contains the ratio $\mu / m_{\tilde Q_3}$.  Due to MSSM RG flow, the squark mass is roughly
\beq
m^2_{\tilde Q_3} \sim \frac{8 \alpha_3}{3} m^2_{\tilde g} \log \frac{\chi_h}{m_{\tilde g}} \ ,
\eeq
where $m_{\tilde g} \gtrsim 400$ GeV is the gluino mass.  As a result, we find that making $\tan \beta$ larger does not significantly increase the hierarchy between the strange and bottom masses.

It is useful to consider a numerical example for the sake of concrete illustration. While this is not intended to exactly reproduce the parameters of the Standard Model fermion sector, it serves to illustrate the significant features of the chiral Higgs model. The parameters in the left column lead to the soft spectrum shown in the right column ($\mu$ and $m_{\tilde \chi}$ are chosen within reason to match $m_Z$):

\begin{table}[h]
\begin{center}
\begin{tabular}{|cc|cc|}
\hline
$\theta_1$ & $\frac{\pi}{4}$ &$m_{\tilde B}$ & $95$ GeV \\
$\theta_2$ & $\frac{\pi}{5}$ & $m_{\tilde W}$ & $190$ GeV \\
$\theta_3$ & $\frac{\pi}{5}$ & $m_{\tilde g}$ & $670$ GeV \\
$\epsilon_l$ & 0.01 &$m_{\tilde \chi}$ & $4.5$ TeV \\
$\epsilon_l'$ & 0.1 & $\mu$ & $200$ GeV \\
$\epsilon_h$ & 0.02 & $\tan \beta$ & 8000 \\
$f$ & 10 TeV & $m_{\tilde t}$ & 700 GeV \\
$\frac{F}{M}$ & 700 TeV &$m_{\tilde \tau_R}$ & 180 GeV \\
$M_1$ & 13 TeV & $m_{\tilde q_{1,2}}$ & $11$ TeV \\
$M_2$ & 19 TeV & $|m_{H_u}|$ & $210$ GeV \\
$M_3$ & 36 TeV & $m_{H_d}$ & $2.4$ TeV \\
\hline
\end{tabular}
\end{center}
\label{tab:num}
\end{table}%

Reasonable  tree-level couplings to the Higgses are given by the Yukawa textures (with a few appropriate $\mathcal{O}(1)$ parameters)
\begin{eqnarray}
Y_u \sim \sin \beta \left(
\begin{array}{ccc}
\epsilon_l & \epsilon_l & \epsilon_h^2 \\
 \epsilon_l & \epsilon_l & \epsilon_h^2 \\
 \epsilon_h^2 & \epsilon_h^2 & 1
\end{array}
\right) \ , \hspace{1cm}
Y_d \sim \cos \beta \left(
\begin{array}{ccc}
 1 & 1.2 & \epsilon_h \\
 1 & 0.3 & \epsilon_h\\
\epsilon_h \epsilon_l' & \epsilon_h \epsilon_l' & \epsilon_l'
\end{array}
\right) \ .
\end{eqnarray}
Ultimately, however, the down quark sector receives mass primarily via radiative corrections. These corrections significantly alter the down-type fermion mass matrix, leading in the infrared to a CKM matrix of the form 
\begin{eqnarray}
|V_{\ rm CKM}| \sim \left(
\begin{array}{ccc}
 0.97 & 0.23 & 0.004 \\
 0.23 & 0.97 & 0.036 \\
 0.004 & 0.037 & 0.99
\end{array}
\right) \ .
\end{eqnarray}

The fermion masses for this specific texture are 
\begin{eqnarray}
\left \{ m_t,m_c,m_u \right \} =\{174 \text{ GeV}, 3.5 \text{ GeV}, 0 \text{ GeV} \} \ ,  \\ \nonumber
\left \{ m_b,m_s,m_d \right \} =\{80 \text{ MeV}, 50 \text{ MeV}, 13 \text{ MeV} \} \ .
\end{eqnarray} 

Here the lightness of the up quark comes from accidental degeneracies in $Y_u$ (as the two-site model cannot generate two quark mass hierarchies without such degeneracies), while the bottom quark comes out anomalously light, as mentioned above.

\bibliography{lit}

\begin{thebibliography}{66}
\expandafter\ifx\csname natexlab\endcsname\relax\def\natexlab#1{#1}\fi
\expandafter\ifx\csname bibnamefont\endcsname\relax
  \def\bibnamefont#1{#1}\fi
\expandafter\ifx\csname bibfnamefont\endcsname\relax
  \def\bibfnamefont#1{#1}\fi
\expandafter\ifx\csname citenamefont\endcsname\relax
  \def\citenamefont#1{#1}\fi
\expandafter\ifx\csname url\endcsname\relax
  \def\url#1{\texttt{#1}}\fi
\expandafter\ifx\csname urlprefix\endcsname\relax\def\urlprefix{URL }\fi
\providecommand{\bibinfo}[2]{#2}
\providecommand{\eprint}[2][]{\url{#2}}

\bibitem[{\citenamefont{Dine and Fischler}(1982{\natexlab{a}})}]{Dine:1981gu}
\bibinfo{author}{\bibfnamefont{M.}~\bibnamefont{Dine}} \bibnamefont{and}
  \bibinfo{author}{\bibfnamefont{W.}~\bibnamefont{Fischler}},
  \emph{\bibinfo{title}{{A Phenomenological Model of Particle Physics Based on
  Supersymmetry}}}, \bibinfo{journal}{Phys. Lett.}
  \textbf{\bibinfo{volume}{B110}}, \bibinfo{pages}{227}
  (\bibinfo{year}{1982}{\natexlab{a}}).

\bibitem[{\citenamefont{Dimopoulos and Raby}(1981)}]{Dimopoulos:1981au}
\bibinfo{author}{\bibfnamefont{S.}~\bibnamefont{Dimopoulos}} \bibnamefont{and}
  \bibinfo{author}{\bibfnamefont{S.}~\bibnamefont{Raby}},
  \emph{\bibinfo{title}{{Supercolor}}}, \bibinfo{journal}{Nucl. Phys.}
  \textbf{\bibinfo{volume}{B192}}, \bibinfo{pages}{353} (\bibinfo{year}{1981}).

\bibitem[{\citenamefont{Alvarez-Gaume et~al.}(1982)\citenamefont{Alvarez-Gaume,
  Claudson, and Wise}}]{AlvarezGaume:1981wy}
\bibinfo{author}{\bibfnamefont{L.}~\bibnamefont{Alvarez-Gaume}},
  \bibinfo{author}{\bibfnamefont{M.}~\bibnamefont{Claudson}}, \bibnamefont{and}
  \bibinfo{author}{\bibfnamefont{M.~B.} \bibnamefont{Wise}},
  \emph{\bibinfo{title}{{Low-Energy Supersymmetry}}}, \bibinfo{journal}{Nucl.
  Phys.} \textbf{\bibinfo{volume}{B207}}, \bibinfo{pages}{96}
  (\bibinfo{year}{1982}).

\bibitem[{\citenamefont{Nappi and Ovrut}(1982)}]{Nappi:1982hm}
\bibinfo{author}{\bibfnamefont{C.~R.} \bibnamefont{Nappi}} \bibnamefont{and}
  \bibinfo{author}{\bibfnamefont{B.~A.} \bibnamefont{Ovrut}},
  \emph{\bibinfo{title}{{Supersymmetric Extension of the $SU(3) x SU(2) x U(1)$
  Model}}}, \bibinfo{journal}{Phys. Lett.} \textbf{\bibinfo{volume}{B113}},
  \bibinfo{pages}{175} (\bibinfo{year}{1982}).

\bibitem[{\citenamefont{Dine and Fischler}(1982{\natexlab{b}})}]{Dine:1982zb}
\bibinfo{author}{\bibfnamefont{M.}~\bibnamefont{Dine}} \bibnamefont{and}
  \bibinfo{author}{\bibfnamefont{W.}~\bibnamefont{Fischler}},
  \emph{\bibinfo{title}{{A Supersymmetric GUT}}}, \bibinfo{journal}{Nucl.
  Phys.} \textbf{\bibinfo{volume}{B204}}, \bibinfo{pages}{346}
  (\bibinfo{year}{1982}{\natexlab{b}}).

\bibitem[{\citenamefont{Dine et~al.}(1995)\citenamefont{Dine, Nelson, and
  Shirman}}]{Dine:1994vc}
\bibinfo{author}{\bibfnamefont{M.}~\bibnamefont{Dine}},
  \bibinfo{author}{\bibfnamefont{A.~E.} \bibnamefont{Nelson}},
  \bibnamefont{and} \bibinfo{author}{\bibfnamefont{Y.}~\bibnamefont{Shirman}},
  \emph{\bibinfo{title}{{Low-energy dynamical supersymmetry breaking
  simplified}}}, \bibinfo{journal}{Phys. Rev.} \textbf{\bibinfo{volume}{D51}},
  \bibinfo{pages}{1362} (\bibinfo{year}{1995}), \eprint{hep-ph/9408384}.

\bibitem[{\citenamefont{Dine et~al.}(1996)\citenamefont{Dine, Nelson, Nir, and
  Shirman}}]{Dine:1995ag}
\bibinfo{author}{\bibfnamefont{M.}~\bibnamefont{Dine}},
  \bibinfo{author}{\bibfnamefont{A.~E.} \bibnamefont{Nelson}},
  \bibinfo{author}{\bibfnamefont{Y.}~\bibnamefont{Nir}}, \bibnamefont{and}
  \bibinfo{author}{\bibfnamefont{Y.}~\bibnamefont{Shirman}},
  \emph{\bibinfo{title}{{New tools for low-energy dynamical supersymmetry
  breaking}}}, \bibinfo{journal}{Phys. Rev.} \textbf{\bibinfo{volume}{D53}},
  \bibinfo{pages}{2658} (\bibinfo{year}{1996}), \eprint{hep-ph/9507378}.

\bibitem[{\citenamefont{Dvali et~al.}(1996)\citenamefont{Dvali, Giudice, and
  Pomarol}}]{Dvali:1996cu}
\bibinfo{author}{\bibfnamefont{G.~R.} \bibnamefont{Dvali}},
  \bibinfo{author}{\bibfnamefont{G.~F.} \bibnamefont{Giudice}},
  \bibnamefont{and} \bibinfo{author}{\bibfnamefont{A.}~\bibnamefont{Pomarol}},
  \emph{\bibinfo{title}{{The $\mu$-Problem in Theories with Gauge-Mediated
  Supersymmetry Breaking}}}, \bibinfo{journal}{Nucl. Phys.}
  \textbf{\bibinfo{volume}{B478}}, \bibinfo{pages}{31} (\bibinfo{year}{1996}),
  \eprint{hep-ph/9603238}.

\bibitem[{\citenamefont{Hall et~al.}(2002)\citenamefont{Hall, Nomura, and
  Pierce}}]{Hall:2002up}
\bibinfo{author}{\bibfnamefont{L.~J.} \bibnamefont{Hall}},
  \bibinfo{author}{\bibfnamefont{Y.}~\bibnamefont{Nomura}}, \bibnamefont{and}
  \bibinfo{author}{\bibfnamefont{A.}~\bibnamefont{Pierce}},
  \emph{\bibinfo{title}{{R symmetry and the mu problem}}},
  \bibinfo{journal}{Phys. Lett.} \textbf{\bibinfo{volume}{B538}},
  \bibinfo{pages}{359} (\bibinfo{year}{2002}), \eprint{hep-ph/0204062}.

\bibitem[{\citenamefont{Giudice et~al.}(2008)\citenamefont{Giudice, Kim, and
  Rattazzi}}]{Giudice:2007ca}
\bibinfo{author}{\bibfnamefont{G.~F.} \bibnamefont{Giudice}},
  \bibinfo{author}{\bibfnamefont{H.~D.} \bibnamefont{Kim}}, \bibnamefont{and}
  \bibinfo{author}{\bibfnamefont{R.}~\bibnamefont{Rattazzi}},
  \emph{\bibinfo{title}{{Natural mu and B mu in gauge mediation}}},
  \bibinfo{journal}{Phys.Lett.} \textbf{\bibinfo{volume}{B660}},
  \bibinfo{pages}{545} (\bibinfo{year}{2008}), \eprint{0711.4448}.

\bibitem[{\citenamefont{Roy and Schmaltz}(2008)}]{Roy:2007nz}
\bibinfo{author}{\bibfnamefont{T.~S.} \bibnamefont{Roy}} \bibnamefont{and}
  \bibinfo{author}{\bibfnamefont{M.}~\bibnamefont{Schmaltz}},
  \emph{\bibinfo{title}{{Hidden solution to the mu/Bmu problem in gauge
  mediation}}}, \bibinfo{journal}{Phys.Rev.} \textbf{\bibinfo{volume}{D77}},
  \bibinfo{pages}{095008} (\bibinfo{year}{2008}), \eprint{0708.3593}.

\bibitem[{\citenamefont{Murayama et~al.}(2008)\citenamefont{Murayama, Nomura,
  and Poland}}]{Murayama:2007ge}
\bibinfo{author}{\bibfnamefont{H.}~\bibnamefont{Murayama}},
  \bibinfo{author}{\bibfnamefont{Y.}~\bibnamefont{Nomura}}, \bibnamefont{and}
  \bibinfo{author}{\bibfnamefont{D.}~\bibnamefont{Poland}},
  \emph{\bibinfo{title}{{More visible effects of the hidden sector}}},
  \bibinfo{journal}{Phys.Rev.} \textbf{\bibinfo{volume}{D77}},
  \bibinfo{pages}{015005} (\bibinfo{year}{2008}), \eprint{0709.0775}.

\bibitem[{\citenamefont{Csaki et~al.}(2009)\citenamefont{Csaki, Falkowski,
  Nomura, and Volansky}}]{Csaki:2008sr}
\bibinfo{author}{\bibfnamefont{C.}~\bibnamefont{Csaki}},
  \bibinfo{author}{\bibfnamefont{A.}~\bibnamefont{Falkowski}},
  \bibinfo{author}{\bibfnamefont{Y.}~\bibnamefont{Nomura}}, \bibnamefont{and}
  \bibinfo{author}{\bibfnamefont{T.}~\bibnamefont{Volansky}},
  \emph{\bibinfo{title}{{New Approach to the $\mu$-$B\mu$ Problem of
  Gauge-Mediated Supersymmetry Breaking}}}, \bibinfo{journal}{Phys. Rev. Lett.}
  \textbf{\bibinfo{volume}{102}}, \bibinfo{pages}{111801}
  (\bibinfo{year}{2009}), \eprint{0809.4492}.

\bibitem[{\citenamefont{Komargodski and Seiberg}(2009)}]{Komargodski:2008ax}
\bibinfo{author}{\bibfnamefont{Z.}~\bibnamefont{Komargodski}} \bibnamefont{and}
  \bibinfo{author}{\bibfnamefont{N.}~\bibnamefont{Seiberg}},
  \emph{\bibinfo{title}{{$\mu$ and General Gauge Mediation}}},
  \bibinfo{journal}{JHEP} \textbf{\bibinfo{volume}{03}}, \bibinfo{pages}{072}
  (\bibinfo{year}{2009}), \eprint{0812.3900}.

\bibitem[{\citenamefont{Green and Weigand}(2009)}]{Green:2009mx}
\bibinfo{author}{\bibfnamefont{D.}~\bibnamefont{Green}} \bibnamefont{and}
  \bibinfo{author}{\bibfnamefont{T.}~\bibnamefont{Weigand}},
  \emph{\bibinfo{title}{{Retrofitting and the mu Problem}}}
  (\bibinfo{year}{2009}), \eprint{0906.0595}.

\bibitem[{\citenamefont{Dine and Kehayias}(2010)}]{Dine:2009swa}
\bibinfo{author}{\bibfnamefont{M.}~\bibnamefont{Dine}} \bibnamefont{and}
  \bibinfo{author}{\bibfnamefont{J.}~\bibnamefont{Kehayias}},
  \emph{\bibinfo{title}{{Discrete R Symmetries and Low Energy Supersymmetry}}},
  \bibinfo{journal}{Phys.Rev.} \textbf{\bibinfo{volume}{D82}},
  \bibinfo{pages}{055014} (\bibinfo{year}{2010}), \eprint{0909.1615}.

\bibitem[{\citenamefont{Evans et~al.}(2011)\citenamefont{Evans, Sudano, and
  Yanagida}}]{Evans:2010ru}
\bibinfo{author}{\bibfnamefont{J.~L.} \bibnamefont{Evans}},
  \bibinfo{author}{\bibfnamefont{M.}~\bibnamefont{Sudano}}, \bibnamefont{and}
  \bibinfo{author}{\bibfnamefont{T.~T.} \bibnamefont{Yanagida}},
  \emph{\bibinfo{title}{{A CP-safe solution of the mu/ Bmu problem of gauge
  mediation}}}, \bibinfo{journal}{Phys.Lett.} \textbf{\bibinfo{volume}{B696}},
  \bibinfo{pages}{348} (\bibinfo{year}{2011}), \eprint{1008.3165}.

\bibitem[{\citenamefont{Schafer-Nameki
  et~al.}(2010{\natexlab{a}})\citenamefont{Schafer-Nameki, Tamarit, and
  Torroba}}]{SchaferNameki:2010mg}
\bibinfo{author}{\bibfnamefont{S.}~\bibnamefont{Schafer-Nameki}},
  \bibinfo{author}{\bibfnamefont{C.}~\bibnamefont{Tamarit}}, \bibnamefont{and}
  \bibinfo{author}{\bibfnamefont{G.}~\bibnamefont{Torroba}},
  \emph{\bibinfo{title}{{Naturalness from runaways in direct mediation}}}
  (\bibinfo{year}{2010}{\natexlab{a}}), \eprint{1011.0001}.

\bibitem[{\citenamefont{Arkani-Hamed
  et~al.}(2001{\natexlab{a}})\citenamefont{Arkani-Hamed, Cohen, and
  Georgi}}]{ArkaniHamed:2001ca}
\bibinfo{author}{\bibfnamefont{N.}~\bibnamefont{Arkani-Hamed}},
  \bibinfo{author}{\bibfnamefont{A.~G.} \bibnamefont{Cohen}}, \bibnamefont{and}
  \bibinfo{author}{\bibfnamefont{H.}~\bibnamefont{Georgi}},
  \emph{\bibinfo{title}{{(De)constructing dimensions}}},
  \bibinfo{journal}{Phys.Rev.Lett.} \textbf{\bibinfo{volume}{86}},
  \bibinfo{pages}{4757} (\bibinfo{year}{2001}{\natexlab{a}}),
  \eprint{hep-th/0104005}.

\bibitem[{\citenamefont{Hill et~al.}(2001)\citenamefont{Hill, Pokorski, and
  Wang}}]{Hill:2000mu}
\bibinfo{author}{\bibfnamefont{C.~T.} \bibnamefont{Hill}},
  \bibinfo{author}{\bibfnamefont{S.}~\bibnamefont{Pokorski}}, \bibnamefont{and}
  \bibinfo{author}{\bibfnamefont{J.}~\bibnamefont{Wang}},
  \emph{\bibinfo{title}{{Gauge invariant effective Lagrangian for Kaluza-Klein
  modes}}}, \bibinfo{journal}{Phys.Rev.} \textbf{\bibinfo{volume}{D64}},
  \bibinfo{pages}{105005} (\bibinfo{year}{2001}), \eprint{hep-th/0104035}.

\bibitem[{\citenamefont{Kaplan et~al.}(2000)\citenamefont{Kaplan, Kribs, and
  Schmaltz}}]{Kaplan:1999ac}
\bibinfo{author}{\bibfnamefont{D.~E.} \bibnamefont{Kaplan}},
  \bibinfo{author}{\bibfnamefont{G.~D.} \bibnamefont{Kribs}}, \bibnamefont{and}
  \bibinfo{author}{\bibfnamefont{M.}~\bibnamefont{Schmaltz}},
  \emph{\bibinfo{title}{{Supersymmetry breaking through transparent extra
  dimensions}}}, \bibinfo{journal}{Phys. Rev.} \textbf{\bibinfo{volume}{D62}},
  \bibinfo{pages}{035010} (\bibinfo{year}{2000}), \eprint{hep-ph/9911293}.

\bibitem[{\citenamefont{Chacko et~al.}(2000)\citenamefont{Chacko, Luty, Nelson,
  and Ponton}}]{Chacko:1999mi}
\bibinfo{author}{\bibfnamefont{Z.}~\bibnamefont{Chacko}},
  \bibinfo{author}{\bibfnamefont{M.~A.} \bibnamefont{Luty}},
  \bibinfo{author}{\bibfnamefont{A.~E.} \bibnamefont{Nelson}},
  \bibnamefont{and} \bibinfo{author}{\bibfnamefont{E.}~\bibnamefont{Ponton}},
  \emph{\bibinfo{title}{{Gaugino mediated supersymmetry breaking}}},
  \bibinfo{journal}{JHEP} \textbf{\bibinfo{volume}{01}}, \bibinfo{pages}{003}
  (\bibinfo{year}{2000}), \eprint{hep-ph/9911323}.

\bibitem[{\citenamefont{Csaki et~al.}(2002{\natexlab{a}})\citenamefont{Csaki,
  Erlich, Grojean, and Kribs}}]{Csaki:2001em}
\bibinfo{author}{\bibfnamefont{C.}~\bibnamefont{Csaki}},
  \bibinfo{author}{\bibfnamefont{J.}~\bibnamefont{Erlich}},
  \bibinfo{author}{\bibfnamefont{C.}~\bibnamefont{Grojean}}, \bibnamefont{and}
  \bibinfo{author}{\bibfnamefont{G.~D.} \bibnamefont{Kribs}},
  \emph{\bibinfo{title}{{4-D constructions of supersymmetric extra dimensions
  and gaugino mediation}}}, \bibinfo{journal}{Phys.Rev.}
  \textbf{\bibinfo{volume}{D65}}, \bibinfo{pages}{015003}
  (\bibinfo{year}{2002}{\natexlab{a}}), \eprint{hep-ph/0106044}.

\bibitem[{\citenamefont{Cheng et~al.}(2001)\citenamefont{Cheng, Kaplan,
  Schmaltz, and Skiba}}]{Cheng:2001an}
\bibinfo{author}{\bibfnamefont{H.}~\bibnamefont{Cheng}},
  \bibinfo{author}{\bibfnamefont{D.}~\bibnamefont{Kaplan}},
  \bibinfo{author}{\bibfnamefont{M.}~\bibnamefont{Schmaltz}}, \bibnamefont{and}
  \bibinfo{author}{\bibfnamefont{W.}~\bibnamefont{Skiba}},
  \emph{\bibinfo{title}{{Deconstructing gaugino mediation}}},
  \bibinfo{journal}{Phys.Lett.} \textbf{\bibinfo{volume}{B515}},
  \bibinfo{pages}{395} (\bibinfo{year}{2001}), \eprint{hep-ph/0106098}.

\bibitem[{\citenamefont{Green et~al.}(2011)\citenamefont{Green, Katz, and
  Komargodski}}]{Green:2010ww}
\bibinfo{author}{\bibfnamefont{D.}~\bibnamefont{Green}},
  \bibinfo{author}{\bibfnamefont{A.}~\bibnamefont{Katz}}, \bibnamefont{and}
  \bibinfo{author}{\bibfnamefont{Z.}~\bibnamefont{Komargodski}},
  \emph{\bibinfo{title}{{Direct Gaugino Mediation}}},
  \bibinfo{journal}{Phys.Rev.Lett.} \textbf{\bibinfo{volume}{106}},
  \bibinfo{pages}{061801} (\bibinfo{year}{2011}), \eprint{1008.2215}.

\bibitem[{\citenamefont{Witten}(2001)}]{Witten:2001bf}
\bibinfo{author}{\bibfnamefont{E.}~\bibnamefont{Witten}},
  \emph{\bibinfo{title}{{Deconstruction, G(2) holonomy, and doublet triplet
  splitting}}}, pp. \bibinfo{pages}{472--491} (\bibinfo{year}{2001}),
  \eprint{hep-ph/0201018}.

\bibitem[{\citenamefont{Dimopoulos and Giudice}(1995)}]{Dimopoulos:1995mi}
\bibinfo{author}{\bibfnamefont{S.}~\bibnamefont{Dimopoulos}} \bibnamefont{and}
  \bibinfo{author}{\bibfnamefont{G.}~\bibnamefont{Giudice}},
  \emph{\bibinfo{title}{{Naturalness constraints in supersymmetric theories
  with nonuniversal soft terms}}}, \bibinfo{journal}{Phys.Lett.}
  \textbf{\bibinfo{volume}{B357}}, \bibinfo{pages}{573} (\bibinfo{year}{1995}),
  \eprint{hep-ph/9507282}.

\bibitem[{\citenamefont{Cohen et~al.}(1996)\citenamefont{Cohen, Kaplan, and
  Nelson}}]{Cohen:1996vb}
\bibinfo{author}{\bibfnamefont{A.~G.} \bibnamefont{Cohen}},
  \bibinfo{author}{\bibfnamefont{D.}~\bibnamefont{Kaplan}}, \bibnamefont{and}
  \bibinfo{author}{\bibfnamefont{A.}~\bibnamefont{Nelson}},
  \emph{\bibinfo{title}{{The More minimal supersymmetric standard model}}},
  \bibinfo{journal}{Phys.Lett.} \textbf{\bibinfo{volume}{B388}},
  \bibinfo{pages}{588} (\bibinfo{year}{1996}), \eprint{hep-ph/9607394}.

\bibitem[{\citenamefont{Arkani-Hamed et~al.}(1998)\citenamefont{Arkani-Hamed,
  Luty, and Terning}}]{ArkaniHamed:1997fq}
\bibinfo{author}{\bibfnamefont{N.}~\bibnamefont{Arkani-Hamed}},
  \bibinfo{author}{\bibfnamefont{M.~A.} \bibnamefont{Luty}}, \bibnamefont{and}
  \bibinfo{author}{\bibfnamefont{J.}~\bibnamefont{Terning}},
  \emph{\bibinfo{title}{{Composite quarks and leptons from dynamical
  supersymmetry breaking without messengers}}}, \bibinfo{journal}{Phys. Rev.}
  \textbf{\bibinfo{volume}{D58}}, \bibinfo{pages}{015004}
  (\bibinfo{year}{1998}), \eprint{hep-ph/9712389}.

\bibitem[{\citenamefont{Luty and Terning}(2000)}]{Luty:1998vr}
\bibinfo{author}{\bibfnamefont{M.~A.} \bibnamefont{Luty}} \bibnamefont{and}
  \bibinfo{author}{\bibfnamefont{J.}~\bibnamefont{Terning}},
  \emph{\bibinfo{title}{{Improved single sector supersymmetry breaking}}},
  \bibinfo{journal}{Phys.Rev.} \textbf{\bibinfo{volume}{D62}},
  \bibinfo{pages}{075006} (\bibinfo{year}{2000}), \eprint{hep-ph/9812290}.

\bibitem[{\citenamefont{Intriligator et~al.}(2006)\citenamefont{Intriligator,
  Seiberg, and Shih}}]{Intriligator:2006dd}
\bibinfo{author}{\bibfnamefont{K.~A.} \bibnamefont{Intriligator}},
  \bibinfo{author}{\bibfnamefont{N.}~\bibnamefont{Seiberg}}, \bibnamefont{and}
  \bibinfo{author}{\bibfnamefont{D.}~\bibnamefont{Shih}},
  \emph{\bibinfo{title}{{Dynamical SUSY breaking in meta-stable vacua}}},
  \bibinfo{journal}{JHEP} \textbf{\bibinfo{volume}{04}}, \bibinfo{pages}{021}
  (\bibinfo{year}{2006}), \eprint{hep-th/0602239}.

\bibitem[{\citenamefont{Franco and Kachru}(2010)}]{Franco:2009wf}
\bibinfo{author}{\bibfnamefont{S.}~\bibnamefont{Franco}} \bibnamefont{and}
  \bibinfo{author}{\bibfnamefont{S.}~\bibnamefont{Kachru}},
  \emph{\bibinfo{title}{{Single-Sector Supersymmetry Breaking in Supersymmetric
  QCD}}}, \bibinfo{journal}{Phys.Rev.} \textbf{\bibinfo{volume}{D81}},
  \bibinfo{pages}{095020} (\bibinfo{year}{2010}), \eprint{0907.2689}.

\bibitem[{\citenamefont{Craig et~al.}(2010)\citenamefont{Craig, Essig, Franco,
  Kachru, and Torroba}}]{Craig:2009hf}
\bibinfo{author}{\bibfnamefont{N.}~\bibnamefont{Craig}},
  \bibinfo{author}{\bibfnamefont{R.}~\bibnamefont{Essig}},
  \bibinfo{author}{\bibfnamefont{S.}~\bibnamefont{Franco}},
  \bibinfo{author}{\bibfnamefont{S.}~\bibnamefont{Kachru}}, \bibnamefont{and}
  \bibinfo{author}{\bibfnamefont{G.}~\bibnamefont{Torroba}},
  \emph{\bibinfo{title}{{Dynamical Supersymmetry Breaking, with Flavor}}},
  \bibinfo{journal}{Phys.Rev.} \textbf{\bibinfo{volume}{D81}},
  \bibinfo{pages}{075015} (\bibinfo{year}{2010}), \eprint{0911.2467}.

\bibitem[{\citenamefont{Behbahani et~al.}(2010)\citenamefont{Behbahani, Craig,
  and Torroba}}]{Behbahani:2010wh}
\bibinfo{author}{\bibfnamefont{S.~R.} \bibnamefont{Behbahani}},
  \bibinfo{author}{\bibfnamefont{N.}~\bibnamefont{Craig}}, \bibnamefont{and}
  \bibinfo{author}{\bibfnamefont{G.}~\bibnamefont{Torroba}},
  \emph{\bibinfo{title}{{Single-sector supersymmetry breaking, chirality, and
  unification}}} (\bibinfo{year}{2010}), \eprint{1009.2088}.

\bibitem[{\citenamefont{Schafer-Nameki
  et~al.}(2010{\natexlab{b}})\citenamefont{Schafer-Nameki, Tamarit, and
  Torroba}}]{SchaferNameki:2010iz}
\bibinfo{author}{\bibfnamefont{S.}~\bibnamefont{Schafer-Nameki}},
  \bibinfo{author}{\bibfnamefont{C.}~\bibnamefont{Tamarit}}, \bibnamefont{and}
  \bibinfo{author}{\bibfnamefont{G.}~\bibnamefont{Torroba}},
  \emph{\bibinfo{title}{{A Hybrid Higgs}}}
  (\bibinfo{year}{2010}{\natexlab{b}}), \eprint{1005.0841}.

\bibitem[{\citenamefont{Hall et~al.}(2004)\citenamefont{Hall, March-Russell,
  Okui, and Tucker-Smith}}]{Hall:2001rz}
\bibinfo{author}{\bibfnamefont{L.~J.} \bibnamefont{Hall}},
  \bibinfo{author}{\bibfnamefont{J.}~\bibnamefont{March-Russell}},
  \bibinfo{author}{\bibfnamefont{T.}~\bibnamefont{Okui}}, \bibnamefont{and}
  \bibinfo{author}{\bibfnamefont{D.}~\bibnamefont{Tucker-Smith}},
  \emph{\bibinfo{title}{{Towards a theory of flavor from orbifold GUTs}}},
  \bibinfo{journal}{JHEP} \textbf{\bibinfo{volume}{0409}}, \bibinfo{pages}{026}
  (\bibinfo{year}{2004}), \eprint{hep-ph/0108161}.

\bibitem[{\citenamefont{Dobrescu and Fox}(2010)}]{Dobrescu:2010mk}
\bibinfo{author}{\bibfnamefont{B.~A.} \bibnamefont{Dobrescu}} \bibnamefont{and}
  \bibinfo{author}{\bibfnamefont{P.~J.} \bibnamefont{Fox}},
  \emph{\bibinfo{title}{{Uplifted supersymmetric Higgs region}}},
  \bibinfo{journal}{Eur. Phys. J. C}  (\bibinfo{year}{2010}),
  \eprint{1001.3147}.

\bibitem[{\citenamefont{Meade et~al.}(2009{\natexlab{a}})\citenamefont{Meade,
  Seiberg, and Shih}}]{Meade:2008wd}
\bibinfo{author}{\bibfnamefont{P.}~\bibnamefont{Meade}},
  \bibinfo{author}{\bibfnamefont{N.}~\bibnamefont{Seiberg}}, \bibnamefont{and}
  \bibinfo{author}{\bibfnamefont{D.}~\bibnamefont{Shih}},
  \emph{\bibinfo{title}{{General Gauge Mediation}}}, \bibinfo{journal}{Prog.
  Theor. Phys. Suppl.} \textbf{\bibinfo{volume}{177}}, \bibinfo{pages}{143}
  (\bibinfo{year}{2009}{\natexlab{a}}), \eprint{0801.3278}.

\bibitem[{\citenamefont{Csaki et~al.}(2002{\natexlab{b}})\citenamefont{Csaki,
  Kribs, and Terning}}]{Csaki:2001qm}
\bibinfo{author}{\bibfnamefont{C.}~\bibnamefont{Csaki}},
  \bibinfo{author}{\bibfnamefont{G.~D.} \bibnamefont{Kribs}}, \bibnamefont{and}
  \bibinfo{author}{\bibfnamefont{J.}~\bibnamefont{Terning}},
  \emph{\bibinfo{title}{{4-D models of Scherk-Schwarz GUT breaking via
  deconstruction}}}, \bibinfo{journal}{Phys.Rev.}
  \textbf{\bibinfo{volume}{D65}}, \bibinfo{pages}{015004}
  (\bibinfo{year}{2002}{\natexlab{b}}), \eprint{hep-ph/0107266}.

\bibitem[{\citenamefont{McGarrie}(2010)}]{McGarrie:2010qr}
\bibinfo{author}{\bibfnamefont{M.}~\bibnamefont{McGarrie}},
  \emph{\bibinfo{title}{{General Gauge Mediation and Deconstruction}}},
  \bibinfo{journal}{JHEP} \textbf{\bibinfo{volume}{11}}, \bibinfo{pages}{152}
  (\bibinfo{year}{2010}), \eprint{1009.0012}.

\bibitem[{\citenamefont{Sudano}(2010)}]{Sudano:2010vt}
\bibinfo{author}{\bibfnamefont{M.}~\bibnamefont{Sudano}},
  \emph{\bibinfo{title}{{General Gaugino Mediation}}} (\bibinfo{year}{2010}),
  \eprint{1009.2086}.

\bibitem[{\citenamefont{Auzzi and Giveon}(2010)}]{Auzzi:2010mb}
\bibinfo{author}{\bibfnamefont{R.}~\bibnamefont{Auzzi}} \bibnamefont{and}
  \bibinfo{author}{\bibfnamefont{A.}~\bibnamefont{Giveon}},
  \emph{\bibinfo{title}{{The Sparticle spectrum in Minimal gaugino-Gauge
  Mediation}}}, \bibinfo{journal}{JHEP} \textbf{\bibinfo{volume}{1010}},
  \bibinfo{pages}{088} (\bibinfo{year}{2010}), \eprint{1009.1714}.

\bibitem[{\citenamefont{Auzzi and Giveon}(2011)}]{Auzzi:2010xc}
\bibinfo{author}{\bibfnamefont{R.}~\bibnamefont{Auzzi}} \bibnamefont{and}
  \bibinfo{author}{\bibfnamefont{A.}~\bibnamefont{Giveon}},
  \emph{\bibinfo{title}{{Superpartner spectrum of minimal gaugino-gauge
  mediation}}}, \bibinfo{journal}{JHEP} \textbf{\bibinfo{volume}{1101}},
  \bibinfo{pages}{003} (\bibinfo{year}{2011}), \eprint{1011.1664}.

\bibitem[{\citenamefont{McGarrie}(2011)}]{McGarrie:2011dc}
\bibinfo{author}{\bibfnamefont{M.}~\bibnamefont{McGarrie}},
  \emph{\bibinfo{title}{{Hybrid Gauge Mediation}}} (\bibinfo{year}{2011}),
  \eprint{1101.5158}.

\bibitem[{\citenamefont{Fan et~al.}(2011)\citenamefont{Fan, Krohn, Mosteiro,
  Thalapillil, and Wang}}]{Fan:2011jc}
\bibinfo{author}{\bibfnamefont{J.}~\bibnamefont{Fan}},
  \bibinfo{author}{\bibfnamefont{D.}~\bibnamefont{Krohn}},
  \bibinfo{author}{\bibfnamefont{P.}~\bibnamefont{Mosteiro}},
  \bibinfo{author}{\bibfnamefont{A.~M.} \bibnamefont{Thalapillil}},
  \bibnamefont{and} \bibinfo{author}{\bibfnamefont{L.-T.} \bibnamefont{Wang}},
  \emph{\bibinfo{title}{{Heavy Squarks at the LHC}}} (\bibinfo{year}{2011}),
  \eprint{1102.0302}.

\bibitem[{\citenamefont{Komargodski and Shih}(2009)}]{Komargodski:2009jf}
\bibinfo{author}{\bibfnamefont{Z.}~\bibnamefont{Komargodski}} \bibnamefont{and}
  \bibinfo{author}{\bibfnamefont{D.}~\bibnamefont{Shih}},
  \emph{\bibinfo{title}{{Notes on SUSY and R-Symmetry Breaking in Wess-Zumino
  Models}}}, \bibinfo{journal}{JHEP} \textbf{\bibinfo{volume}{04}},
  \bibinfo{pages}{093} (\bibinfo{year}{2009}), \eprint{0902.0030}.

\bibitem[{\citenamefont{Kitano et~al.}(2007)\citenamefont{Kitano, Ooguri, and
  Ookouchi}}]{Kitano:2006xg}
\bibinfo{author}{\bibfnamefont{R.}~\bibnamefont{Kitano}},
  \bibinfo{author}{\bibfnamefont{H.}~\bibnamefont{Ooguri}}, \bibnamefont{and}
  \bibinfo{author}{\bibfnamefont{Y.}~\bibnamefont{Ookouchi}},
  \emph{\bibinfo{title}{{Direct mediation of meta-stable supersymmetry
  breaking}}}, \bibinfo{journal}{Phys. Rev.} \textbf{\bibinfo{volume}{D75}},
  \bibinfo{pages}{045022} (\bibinfo{year}{2007}), \eprint{hep-ph/0612139}.

\bibitem[{\citenamefont{Giveon et~al.}(2009)\citenamefont{Giveon, Katz, and
  Komargodski}}]{Giveon:2009yu}
\bibinfo{author}{\bibfnamefont{A.}~\bibnamefont{Giveon}},
  \bibinfo{author}{\bibfnamefont{A.}~\bibnamefont{Katz}}, \bibnamefont{and}
  \bibinfo{author}{\bibfnamefont{Z.}~\bibnamefont{Komargodski}},
  \emph{\bibinfo{title}{{Uplifted Metastable Vacua and Gauge Mediation in
  SQCD}}} (\bibinfo{year}{2009}), \eprint{0905.3387}.

\bibitem[{\citenamefont{Koschade et~al.}(2010)\citenamefont{Koschade, McGarrie,
  and Thomas}}]{Koschade:2009qu}
\bibinfo{author}{\bibfnamefont{D.}~\bibnamefont{Koschade}},
  \bibinfo{author}{\bibfnamefont{M.}~\bibnamefont{McGarrie}}, \bibnamefont{and}
  \bibinfo{author}{\bibfnamefont{S.}~\bibnamefont{Thomas}},
  \emph{\bibinfo{title}{{Direct Mediation and Metastable Supersymmetry Breaking
  for SO(10)}}}, \bibinfo{journal}{JHEP} \textbf{\bibinfo{volume}{02}},
  \bibinfo{pages}{100} (\bibinfo{year}{2010}), \eprint{0909.0233}.

\bibitem[{\citenamefont{Barnard}(2010)}]{Barnard:2009ir}
\bibinfo{author}{\bibfnamefont{J.}~\bibnamefont{Barnard}},
  \emph{\bibinfo{title}{{Tree Level Metastability and Gauge Mediation in Baryon
  Deformed SQCD}}}, \bibinfo{journal}{JHEP} \textbf{\bibinfo{volume}{02}},
  \bibinfo{pages}{035} (\bibinfo{year}{2010}), \eprint{0910.4047}.

\bibitem[{\citenamefont{Maru}(2010)}]{Maru:2010yx}
\bibinfo{author}{\bibfnamefont{N.}~\bibnamefont{Maru}},
  \emph{\bibinfo{title}{{Direct Gauge Mediation of Uplifted Metastable
  Supersymmetry Breaking in Supergravity}}}, \bibinfo{journal}{Phys. Rev.}
  \textbf{\bibinfo{volume}{D82}}, \bibinfo{pages}{075015}
  (\bibinfo{year}{2010}), \eprint{1008.1440}.

\bibitem[{\citenamefont{Curtin and Tsai}(2010)}]{Curtin:2010ku}
\bibinfo{author}{\bibfnamefont{D.}~\bibnamefont{Curtin}} \bibnamefont{and}
  \bibinfo{author}{\bibfnamefont{Y.}~\bibnamefont{Tsai}},
  \emph{\bibinfo{title}{{Singlet-Stabilized Minimal Gauge Mediation}}}
  (\bibinfo{year}{2010}), \eprint{1011.2766}.

\bibitem[{\citenamefont{Giveon et~al.}(2008)\citenamefont{Giveon, Katz,
  Komargodski, and Shih}}]{Giveon:2008ne}
\bibinfo{author}{\bibfnamefont{A.}~\bibnamefont{Giveon}},
  \bibinfo{author}{\bibfnamefont{A.}~\bibnamefont{Katz}},
  \bibinfo{author}{\bibfnamefont{Z.}~\bibnamefont{Komargodski}},
  \bibnamefont{and} \bibinfo{author}{\bibfnamefont{D.}~\bibnamefont{Shih}},
  \emph{\bibinfo{title}{{Dynamical SUSY and R-symmetry breaking in SQCD with
  massive and massless flavors}}}, \bibinfo{journal}{JHEP}
  \textbf{\bibinfo{volume}{10}}, \bibinfo{pages}{092} (\bibinfo{year}{2008}),
  \eprint{0808.2901}.

\bibitem[{\citenamefont{{De Simone} et~al.}(2008)\citenamefont{{De Simone},
  Fan, Schmaltz, and Skiba}}]{DeSimone:2008gm}
\bibinfo{author}{\bibfnamefont{A.}~\bibnamefont{{De Simone}}},
  \bibinfo{author}{\bibfnamefont{J.}~\bibnamefont{Fan}},
  \bibinfo{author}{\bibfnamefont{M.}~\bibnamefont{Schmaltz}}, \bibnamefont{and}
  \bibinfo{author}{\bibfnamefont{W.}~\bibnamefont{Skiba}},
  \emph{\bibinfo{title}{{Low-scale gaugino mediation, lots of leptons at the
  LHC}}}, \bibinfo{journal}{Phys. Rev.} \textbf{\bibinfo{volume}{D78}},
  \bibinfo{pages}{095010} (\bibinfo{year}{2008}), \eprint{0808.2052}.

\bibitem[{\citenamefont{Maloney et~al.}(2006)\citenamefont{Maloney, Pierce, and
  Wacker}}]{Maloney:2004rc}
\bibinfo{author}{\bibfnamefont{A.}~\bibnamefont{Maloney}},
  \bibinfo{author}{\bibfnamefont{A.}~\bibnamefont{Pierce}}, \bibnamefont{and}
  \bibinfo{author}{\bibfnamefont{J.~G.} \bibnamefont{Wacker}},
  \emph{\bibinfo{title}{{D-terms, unification, and the Higgs mass}}},
  \bibinfo{journal}{JHEP} \textbf{\bibinfo{volume}{0606}}, \bibinfo{pages}{034}
  (\bibinfo{year}{2006}), \eprint{hep-ph/0409127}.

\bibitem[{\citenamefont{Delgado et~al.}(2000)\citenamefont{Delgado, Pomarol,
  and Quiros}}]{Delgado:1999sv}
\bibinfo{author}{\bibfnamefont{A.}~\bibnamefont{Delgado}},
  \bibinfo{author}{\bibfnamefont{A.}~\bibnamefont{Pomarol}}, \bibnamefont{and}
  \bibinfo{author}{\bibfnamefont{M.}~\bibnamefont{Quiros}},
  \emph{\bibinfo{title}{{Electroweak and flavor physics in extensions of the
  standard model with large extra dimensions}}}, \bibinfo{journal}{JHEP}
  \textbf{\bibinfo{volume}{01}}, \bibinfo{pages}{030} (\bibinfo{year}{2000}),
  \eprint{hep-ph/9911252}.

\bibitem[{\citenamefont{Bona et~al.}(2008)}]{Bona:2007vi}
\bibinfo{author}{\bibfnamefont{M.}~\bibnamefont{Bona}} \bibnamefont{et~al.}
  (\bibinfo{collaboration}{UTfit}), \emph{\bibinfo{title}{{Model-independent
  constraints on $\Delta$ F=2 operators and the scale of new physics}}},
  \bibinfo{journal}{JHEP} \textbf{\bibinfo{volume}{03}}, \bibinfo{pages}{049}
  (\bibinfo{year}{2008}), \eprint{0707.0636}.

\bibitem[{\citenamefont{Bertolini et~al.}(1991)\citenamefont{Bertolini,
  Borzumati, Masiero, and Ridolfi}}]{Bertolini:1990if}
\bibinfo{author}{\bibfnamefont{S.}~\bibnamefont{Bertolini}},
  \bibinfo{author}{\bibfnamefont{F.}~\bibnamefont{Borzumati}},
  \bibinfo{author}{\bibfnamefont{A.}~\bibnamefont{Masiero}}, \bibnamefont{and}
  \bibinfo{author}{\bibfnamefont{G.}~\bibnamefont{Ridolfi}},
  \emph{\bibinfo{title}{{Effects of supergravity induced electroweak breaking
  on rare $B$ decays and mixings}}}, \bibinfo{journal}{Nucl.Phys.}
  \textbf{\bibinfo{volume}{B353}}, \bibinfo{pages}{591} (\bibinfo{year}{1991}).

\bibitem[{\citenamefont{Giudice et~al.}(2009)\citenamefont{Giudice, Nardecchia,
  and Romanino}}]{Giudice:2008uk}
\bibinfo{author}{\bibfnamefont{G.~F.} \bibnamefont{Giudice}},
  \bibinfo{author}{\bibfnamefont{M.}~\bibnamefont{Nardecchia}},
  \bibnamefont{and} \bibinfo{author}{\bibfnamefont{A.}~\bibnamefont{Romanino}},
  \emph{\bibinfo{title}{{Hierarchical Soft Terms and Flavor Physics}}},
  \bibinfo{journal}{Nucl.Phys.} \textbf{\bibinfo{volume}{B813}},
  \bibinfo{pages}{156} (\bibinfo{year}{2009}), \eprint{0812.3610}.

\bibitem[{\citenamefont{Nakamura et~al.}(2010)}]{Nakamura:2010zzi}
\bibinfo{author}{\bibfnamefont{K.}~\bibnamefont{Nakamura}} \bibnamefont{et~al.}
  (\bibinfo{collaboration}{Particle Data Group Collaboration}),
  \emph{\bibinfo{title}{{Review of particle physics}}}, \bibinfo{journal}{J.\
  Phys.\ G} \textbf{\bibinfo{volume}{G37}}, \bibinfo{pages}{075021}
  (\bibinfo{year}{2010}).

\bibitem[{\citenamefont{{The Heavy Flavor Averaging Group}
  et~al.}(2010)}]{TheHeavyFlavorAveragingGroup:2010qj}
\bibinfo{author}{\bibnamefont{{The Heavy Flavor Averaging Group}}}
  \bibnamefont{et~al.}, \emph{\bibinfo{title}{{Averages of b-hadron, c-hadron,
  and tau-lepton Properties}}} (\bibinfo{year}{2010}), \eprint{1010.1589}.

\bibitem[{\citenamefont{Buras and Guadagnoli}(2008)}]{Buras:2008nn}
\bibinfo{author}{\bibfnamefont{A.~J.} \bibnamefont{Buras}} \bibnamefont{and}
  \bibinfo{author}{\bibfnamefont{D.}~\bibnamefont{Guadagnoli}},
  \emph{\bibinfo{title}{{Correlations among new CP violating effects in Delta F
  = 2 observables}}}, \bibinfo{journal}{Phys. Rev.}
  \textbf{\bibinfo{volume}{D78}}, \bibinfo{pages}{033005}
  (\bibinfo{year}{2008}), \eprint{0805.3887}.

\bibitem[{\citenamefont{Arkani-Hamed and Murayama}(1997)}]{ArkaniHamed:1997ab}
\bibinfo{author}{\bibfnamefont{N.}~\bibnamefont{Arkani-Hamed}}
  \bibnamefont{and} \bibinfo{author}{\bibfnamefont{H.}~\bibnamefont{Murayama}},
  \emph{\bibinfo{title}{{Can the supersymmetric flavor problem decouple?}}},
  \bibinfo{journal}{Phys.Rev.} \textbf{\bibinfo{volume}{D56}},
  \bibinfo{pages}{6733} (\bibinfo{year}{1997}), \eprint{hep-ph/9703259}.

\bibitem[{\citenamefont{Schmaltz and Skiba}(2000)}]{Schmaltz:2000gy}
\bibinfo{author}{\bibfnamefont{M.}~\bibnamefont{Schmaltz}} \bibnamefont{and}
  \bibinfo{author}{\bibfnamefont{W.}~\bibnamefont{Skiba}},
  \emph{\bibinfo{title}{{Minimal gaugino mediation}}},
  \bibinfo{journal}{Phys.Rev.} \textbf{\bibinfo{volume}{D62}},
  \bibinfo{pages}{095005} (\bibinfo{year}{2000}), \eprint{hep-ph/0001172}.

\bibitem[{\citenamefont{Meade et~al.}(2009{\natexlab{b}})\citenamefont{Meade,
  Reece, and Shih}}]{Meade:2009qv}
\bibinfo{author}{\bibfnamefont{P.}~\bibnamefont{Meade}},
  \bibinfo{author}{\bibfnamefont{M.}~\bibnamefont{Reece}}, \bibnamefont{and}
  \bibinfo{author}{\bibfnamefont{D.}~\bibnamefont{Shih}},
  \emph{\bibinfo{title}{{Prompt Decays of General Neutralino NLSPs at the
  Tevatron}}} (\bibinfo{year}{2009}{\natexlab{b}}), \eprint{0911.4130}.

\bibitem[{\citenamefont{Arkani-Hamed
  et~al.}(2001{\natexlab{b}})\citenamefont{Arkani-Hamed, Cohen, and
  Georgi}}]{ArkaniHamed:2001vr}
\bibinfo{author}{\bibfnamefont{N.}~\bibnamefont{Arkani-Hamed}},
  \bibinfo{author}{\bibfnamefont{A.~G.} \bibnamefont{Cohen}}, \bibnamefont{and}
  \bibinfo{author}{\bibfnamefont{H.}~\bibnamefont{Georgi}},
  \emph{\bibinfo{title}{{Accelerated unification}}}
  (\bibinfo{year}{2001}{\natexlab{b}}), \eprint{hep-th/0108089}.

\end{thebibliography}
\bibliographystyle{apsper}

\end{document}